\documentclass[twocolumn]{aastex631}
\usepackage{bm}
\usepackage{graphicx}
\usepackage{txfonts}
\usepackage[utf8]{inputenc}
\usepackage{natbib}
\usepackage{float}
\usepackage{mathrsfs}
\usepackage{hyperref}
\usepackage{esint}

\received{2023 December 11}
\revised{2024 January 30}
\accepted{2024 May 28}
\published{}
\submitjournal{\apj}

\setwatermarkfontsize{2.5in}

\shorttitle{Refinement of global CMF and IMF constrained by remote-sensing and in-situ observations}
\shortauthors{Shi et al.}

\graphicspath{{./}{figures/}}

\begin{document}

\title{Refinement of global coronal and interplanetary magnetic field extrapolations constrained \\ by remote-sensing and in-situ observations at the solar minimum}

\author[0000-0001-7397-455X]{Guanglu Shi}
\affiliation{Key Laboratory of Dark Matter and Space Astronomy, Purple Mountain Observatory, 
\\ Chinese Academy of Sciences, Nanjing 210023, People's Republic of China}
\affiliation{School of Astronomy and Space Science, University of Science and Technology of China, 
\\ Hefei, Anhui 230026, People's Republic of China}

\author[0000-0003-4655-6939]{Li Feng}
\affiliation{Key Laboratory of Dark Matter and Space Astronomy, Purple Mountain Observatory, 
\\ Chinese Academy of Sciences, Nanjing 210023, People's Republic of China}
\affiliation{School of Astronomy and Space Science, University of Science and Technology of China, 
\\ Hefei, Anhui 230026, People's Republic of China}
\email{lfeng@pmo.ac.cn}

\author[0000-0001-8402-9748]{Beili Ying}
\affiliation{Key Laboratory of Dark Matter and Space Astronomy, Purple Mountain Observatory, 
\\ Chinese Academy of Sciences, Nanjing 210023, People's Republic of China}
\affiliation{School of Astronomy and Space Science, University of Science and Technology of China, 
\\ Hefei, Anhui 230026, People's Republic of China}

\author[0000-0003-2694-2875]{Shuting Li}
\affiliation{Key Laboratory of Dark Matter and Space Astronomy, Purple Mountain Observatory, 
\\ Chinese Academy of Sciences, Nanjing 210023, People's Republic of China}
\affiliation{School of Astronomy and Space Science, University of Science and Technology of China,
\\ Hefei, Anhui 230026, People's Republic of China}

\author[0000-0001-9979-4178]{Weiqun Gan}
\affiliation{Key Laboratory of Dark Matter and Space Astronomy, Purple Mountain Observatory, 
\\ Chinese Academy of Sciences, Nanjing 210023, People's Republic of China}
\affiliation{University of Chinese Academy of Sciences, Nanjing 211135, People's Republic of China}

\begin{abstract}

Solar magnetic fields are closely related to various physical phenomena on the sun, which can be extrapolated with different models from photospheric magnetograms. However, the Open Flux Problem (OFP), the underestimation of the magnetic field derived from the extrapolated model, is still unsolved. To minimize the impact of the OFP, we propose three evaluation parameters to quantitatively evaluate magnetic field models and determine the optimal free parameters in the models by constraining the coronal magnetic fields (CMFs) and the interplanetary magnetic fields (IMFs) with real observations. Although the OFP still exists, we find that magnetic field lines traced from the coronal models effectively capture the intricate topological configurations observed in the corona, including streamers and plumes. The OFP is lessened by using the HMI synoptic map instead of the GONG daily synoptic maps, and the PFSS+PFCS model instead of the CSSS model. For Carrington Rotation (CR) 2231 at the solar minimum, we suggest that the optimal parameters for the PFSS+PFCS model are $R_{\rm ss} = 2.2-2.5\ R_\sun$ and $R_{\rm scs} = 10.5-14.0\ R_\sun$, as well as for the CSSS model are $R_{\rm cs} = 2.0 - 2.4\ R_\sun$, $R_{\rm ss} = 11.0 - 14.7\ R_\sun$ and $a = 1.0\ R_\sun$. Despite the IMFs at $1\ {\rm AU}$ being consistent with the measurements by artificially increasing the polar magnetic fields, the IMFs near the sun are still underestimated. The OFP might be advanced by improving the accuracy of both the weak magnetic fields and polar magnetic fields, especially considering magnetic activities arising from interplanetary physical processes.

\end{abstract}



\section{Introduction} \label{sec:intro}

The solar magnetic fields play a crucial role in various solar activities. Open magnetic field lines extend from the solar surface to the heliosphere, forming the interplanetary magnetic fields (IMFs). In interplanetary space, plasma beta is defined as the ratio of thermal pressure to magnetic pressure, $\beta \gg 1$, and flows dominate the magnetic fields. Closed magnetic fields are mainly distributed in the low corona region, $\beta \ll 1$, and the magnetic fields dominate movements of the plasma \citep{Priest1991, aschwanden2006}. In models \citep{Mackay2012, Priest2014} and observations \citep{Panesar2021ApJ, Wang2022ApJ939, Perrone2022A&A, Badman2023JGRA, Bale2023Natur} of the coronal magnetic fields (CMFs), the open magnetic fields originate from the Coronal Holes (CHs) connecting to the dark regions on the solar disk observed in the Extreme Ultraviolet (EUV) and X-ray images \citep{Bohlin1977, Zirker1977} contributing to the heliospheric flux, which is related to the origin of the fast solar wind. The closed magnetic fields usually dominate in the active region and the quiet sun.

Quantifying the splitting of spectral lines caused by the Zeeman effect is an essential method to measure the magnetic fields. Although direct measurements of weak CMFs are challenging, some new diagnostic techniques have been developed \citep{Yang2020, Chen2020NatAs, Wei2021ApJ, Tan2022RAA, Trujillo2022, Landi2022, Bemporad2023ApJ, Chen2023MNRAS}. The IMFs can be indirectly measured through the detection of cosmic-ray sun shadow, as demonstrated by the Tibet AS-$\gamma$ \citep{Amenomori2013PhRvL, Amenomori2018PhRvL}, IceCube \citep{Aartsen2017JInst, Aartsen2021PhRvD}, and Large High Altitude Air Shower Observatory \citep[LHAASO, ][]{Cao2019arXiv, Nan2022icrc, Jia2022ChPhC, Li2023ExA} experiments.

Lots of theoretical models have been constructed to investigate the solar magnetic fields, such as the Potential Field Source Surface \citep[PFSS,][]{Altschuler1969, Schatten1969} model, the Potential Field Current Sheet \citep[PFCS,][]{Schatten1972} model, the Horizontal Current Source Surface \citep[HCSS,][]{Zhao1993SoPh} model, the Horizontal Current–Current Sheet \citep[HCCS,][]{Zhao1994SoPh} model, the Current Sheet Source Surface \citep[CSSS,][]{Zhao1995, Zhao2002} model, the Nonlinear Force–Free Field \citep[NLFFF,][]{Wiegelmann2007, Contopoulos2011, Yeates2018ssr} model, and many other Magnetohydrodynamic \citep[MHD,][]{Linker1999, Odstrcil2003, Sokolov2013ApJ, Holst2014ApJ, Pomoell2018, Riley2019, Holst2019, Holst2022, Sokolov2022ApJ} models. Comparisons among these models and the optimal free parameters have been widely studied \citep{Schussler2006, Koskela2019, Wagner2022A&A}. Many studies have confirmed that the IMFs predicted by models consistently underestimate their strength compared to in-situ measurements obtained near the Earth \citep{Svalgaard1978, Wang1995, Ulrich2009, Riley2012, Wallace2019}, called the Open Flux Problem \citep[OFP, ][]{Linker2017, Riley2019a, Badman2021A&A, Wang2022ApJ, Arge2023arXiv, Yoshida2023ApJ}. At the meantime, observations from Ulysses indicated that the radial component of the IMFs is independent of latitude \citep{Balogh1995, Smith1995}. In recent years, with the launch of spacecraft approaching the sun and navigating outside the ecliptic plane, such as the Parker Solar Probe \citep[PSP,][]{Fox2018}, and the Solar Orbiter \citep[SolO,][]{Muller2020}, multi-perspective observations enhance our ability to stereoscopic image the sun and verify the performance of models. Although there is good consistency between predictions by models and observations in qualitative comparisons, such as topological configurations and polarities, the OFP still exists in the interplanetary space in quantitative comparisons \citep{Badman2020, Riley2021, Badman2021, Song2023}. 

In this work, we extrapolate the global CMFs and IMFs via different models during the Carrington Rotation (CR) 2231 from May 21 to June 18, 2020, at the solar minimum. To better constrain different extrapolation models, we use both remote-sensing and in-situ observations from multi-perspective spacecraft. We explore variations of CMFs and IMFs with longitude, latitude, and distance, then try to investigate the OFP. Section \ref{sec:obs} describes available observations used in this work. The PFSS, PFCS, CSSS, and Parker spiral models are introduced in Section \ref{sec:mag_mod}. Section \ref{sec:results} shows our results: (1) topological configurations and heliospheric current sheets (HCSs) of CMFs; (2) the establishment of three evaluation parameters for quantitative comparisons between models and observations. Discussion and conclusions are given in Section \ref{sec:conclusions}.

\section{Available observations} \label{sec:obs}

\subsection{Remote-sensing observations}

\subsubsection{Photospheric magnetogram}

Two types of products are available in photospheric magnetograms: diachronic synoptic maps and synchronic synoptic maps \citep{Linker2017}. We select Helioseismic and Magnetic Imager \citep[HMI,][]{Schou2012} synoptic map and Global Oscillation Network Group (GONG) daily synoptic maps as representatives between these two types for comparisons, respectively.

\begin{figure*}[htb!]
\begin{center}
\includegraphics[height=0.19\textheight]{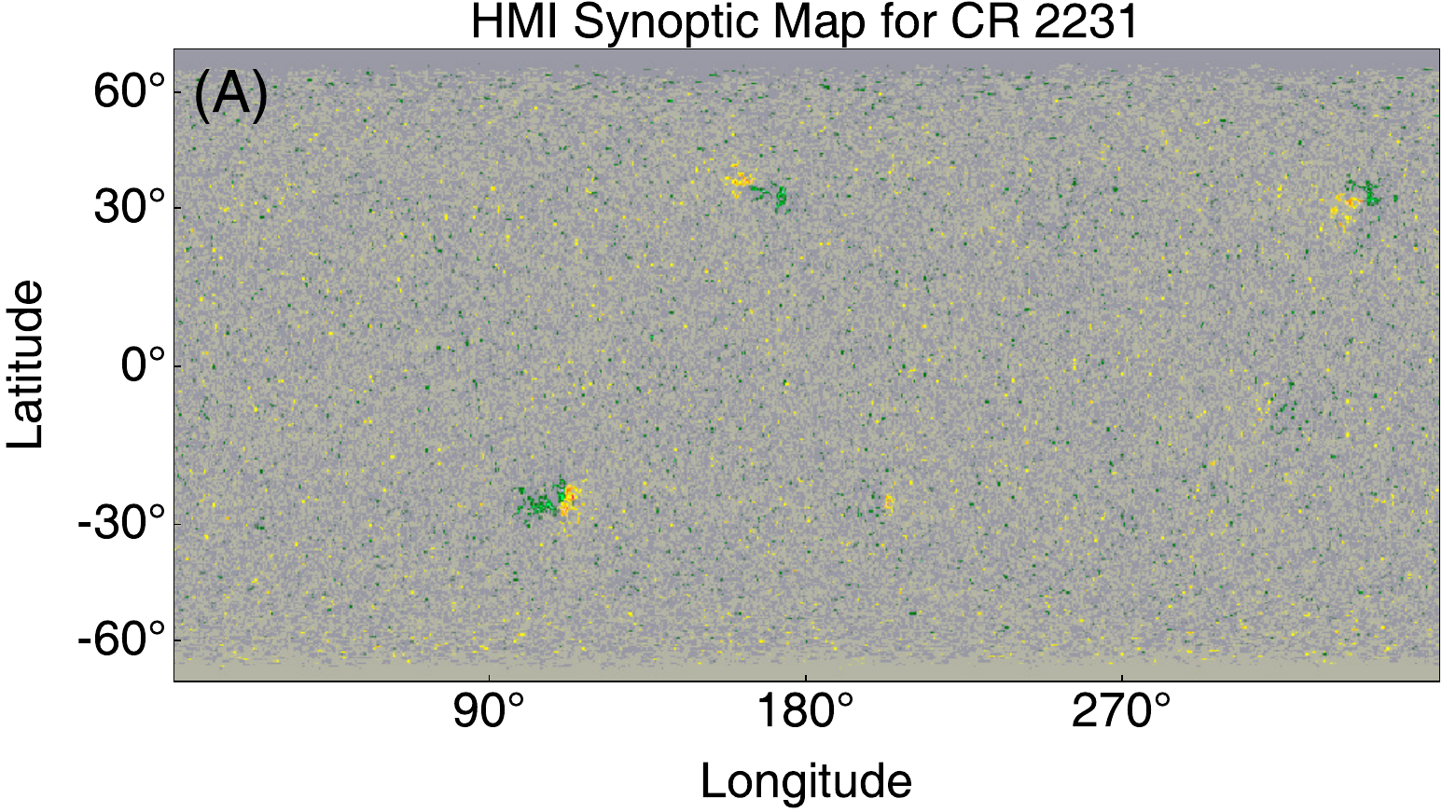}
\includegraphics[height=0.19\textheight]{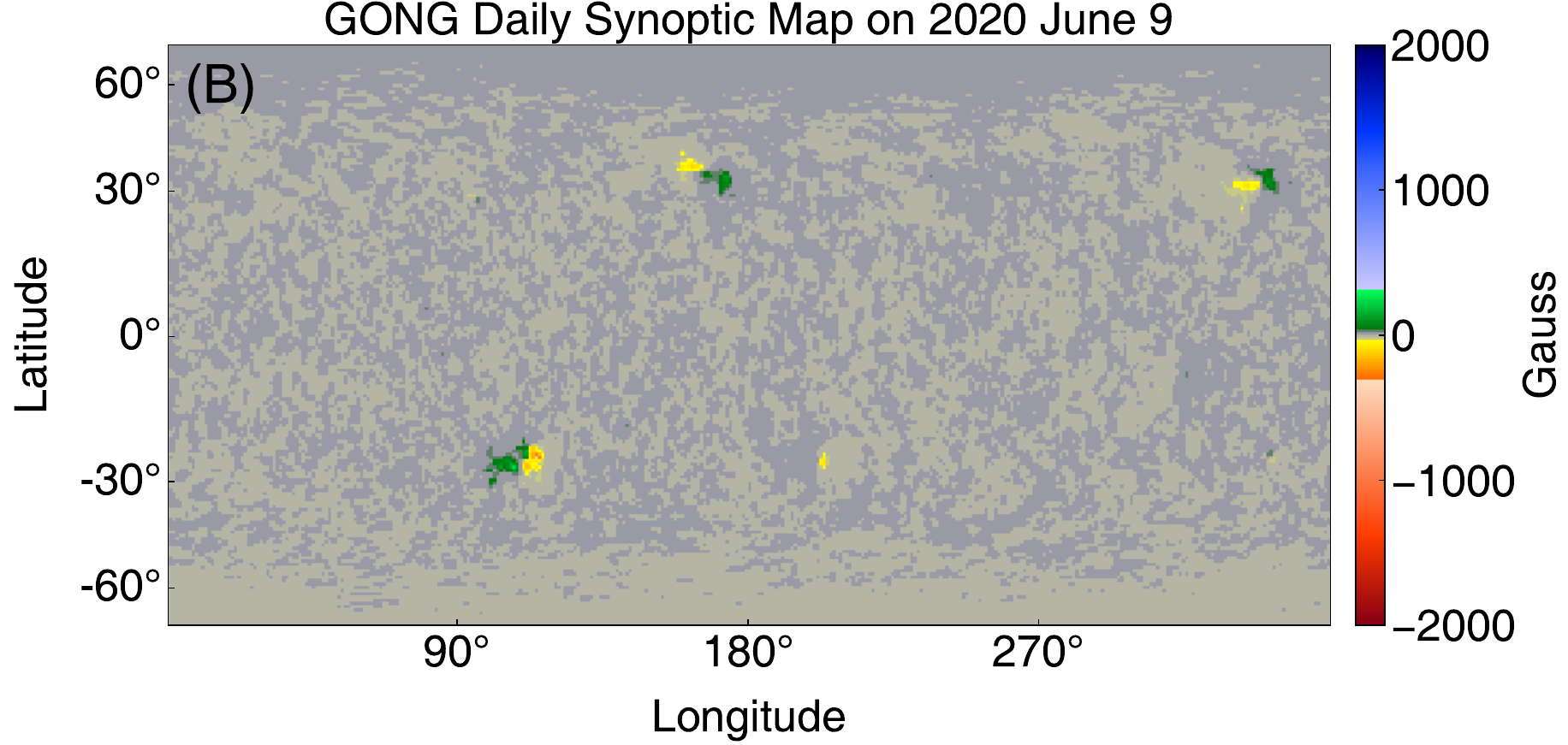}
\caption{(A) HMI synoptic map for CR 2231. (B) Zero-point corrected GONG daily synoptic map updated on June 9, 2020. \label{fig:hmi_gong}}
\end{center}
\end{figure*}

HMI is one of three payloads onboard the Solar Dynamics Observatory \citep[SDO,][]{Hoeksema2014}. HMI synoptic maps are built up diachronically. Specifically, they are spliced from the Line-of-Sight (LoS) magnetograms of daily observations on the solar full-disk with a cadence of 720s, whose dimension is $3600$ (longitude, $\phi$) $\times$ $1440$ (sine-latitude, $\sin \theta$). Due to the orbit of the SDO near the ecliptic, normal HMI synoptic maps miss polar observations. In this work, a method \citep{Sun2018} of spatial-temporal interpolation is used to correct polar magnetic fields in the HMI synoptic map, shown in Figure \ref{fig:hmi_gong} (A). Considering that the dimensions of the HMI synoptic map are too large for practical calculations of magnetic field models, we rescale it to $720 \times 360$.

GONG is a community-based program to study solar internal structures and dynamics using helioseismology. GONG daily synoptic maps are combined from solar full-disk magnetograms observed by a six-station network. The spatial resolution of the GONG synoptic map is lower than the HMI, and the dimension is $360 \times 180$. Full-disk magnetograms generated each hour at each station can update the synoptic maps in quasi-real time. The measurement accuracy of the GONG synoptic map is impacted by the zero-point error due to instrument effects. The noise level is about $3\ {\rm G}$ per pixel, while the zero-point error may be incorrect by as much as $10\ {\rm G}$. Hence, zero-point corrected GONG daily synoptic maps \citep{Clark2003} totaling 32 maps from May 20 to June 20, 2020, are used as boundary conditions. Figure \ref{fig:hmi_gong} (B) shows one GONG daily synoptic map updated on June 9, 2020, as an example.

\subsubsection{EUV and coronagraph observations}

\begin{figure*}[htb!]
\begin{center}
\includegraphics[width=0.95\textwidth]{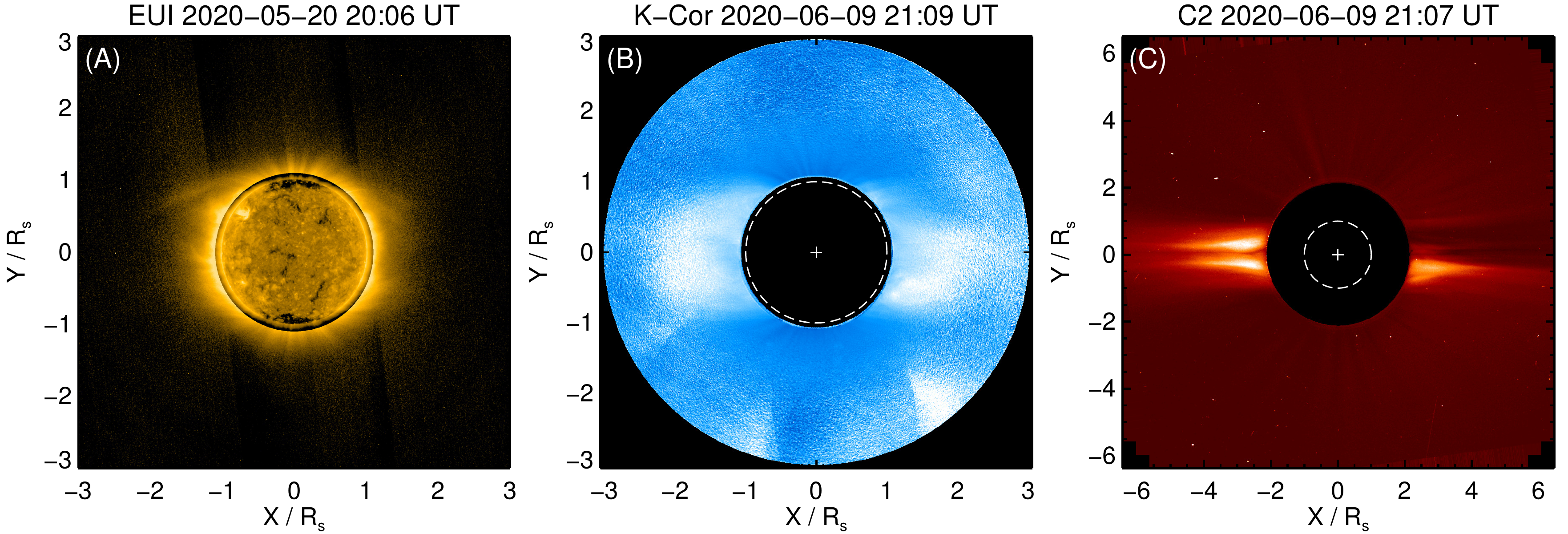}
\caption{EUV and VL coronagraphic images. (A) Radial filtered EUV image observed by the EUI. (B) \& (C) VL coronagraphic images observed by the K-Cor and LASCO C2, respectively. Dashed lines and cross symbols indicate the edge and the center of the sun, respectively. \label{fig:euv_wl}}
\end{center}
\end{figure*}

Figure \ref{fig:euv_wl} shows the images with different Field-of-Views (FoVs) observed by the Extreme Ultraviolet Imager \citep[EUI,][]{Rochus2020} onboard the SolO, K-Coronagraph \citep[K-Cor,][]{Burkepile2013} at the Mauna Loa Solar Observatory (MLSO), and Large Angle and Spectrometric Coronagraph \citep[LASCO,][]{Brueckner1995} onboard the Solar and Heliospheric Observatory \citep[SOHO,][]{Domingo1995}, which can study the global CMF configurations extend from the solar surface to the corona.

The EUI consists of three telescopes, a Full Sun Imager (FSI), and two High Resolution Imagers (HRIs), which can provide large FoV and high dynamic range observations of the sun to study the formation and evolution of solar activities. Figure \ref{fig:euv_wl} (A) shows an image with a FoV of $3.0\ R_\sun$ observed by the FSI 174 \AA. The effective exposure time was 10 s at 20:06 UT on May 20, 2020, resulting in overexposure of the solar disk, which is replaced by the image at 20:08 UT with an effective exposure time of 5 s. Meanwhile, the radial filtering method is employed to enhance coronal features. The internal structures of the streamers distributed along the equator and the plumes rooted in polar regions are visible.

Figure \ref{fig:euv_wl} (B) \& (C) show Visible Light (VL) images of the K-Cor and the LASCO C2 taken on June 9, 2020, providing continuous observations from the low to high corona. The FoV of the K-Cor is $1.05 - 3.0\ R_\sun$, and the LASCO C2 is $2.0 - 6.0\ R_\sun$. The image of the LASCO C2 is subtracted by the monthly-minimum background to eliminate the influence of the F-corona and instrument stray light. The dashed line indicates the solar limb, and the cross symbol denotes the center of the sun in each panel. Streamers are distributed around the equator and become open until $4.0\ R_\sun$. At polar regions, plumes related to the open magnetic fields extend to interplanetary space.

\begin{figure*}[htb!]
\begin{center}
\includegraphics[width=0.9\textwidth]{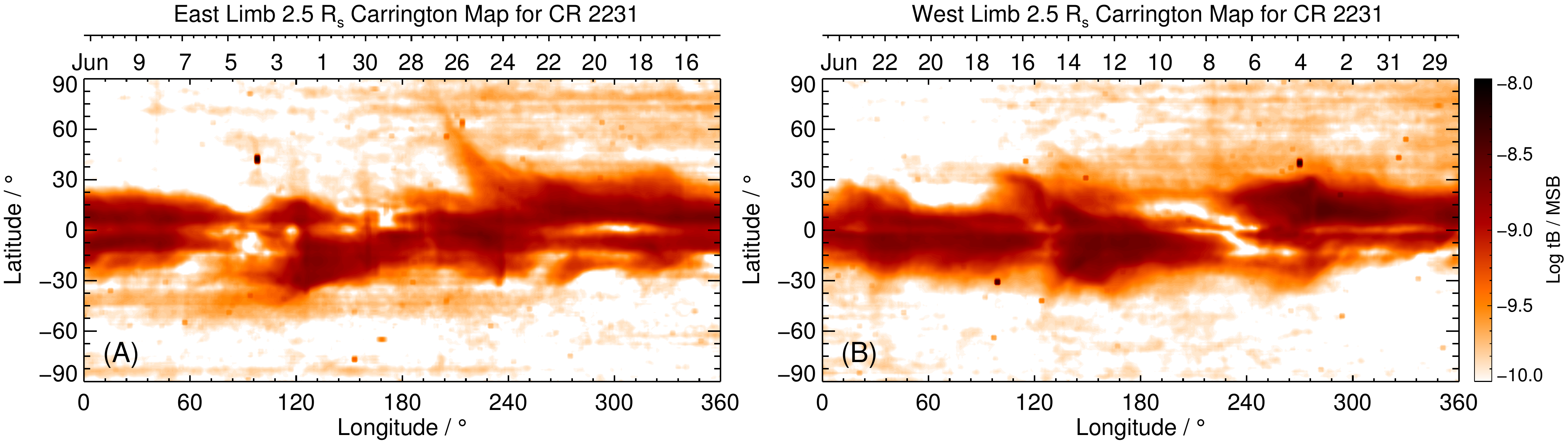}
\caption{East (A) and west (B) limb Carrington maps spliced by images of the LASCO C2 at $2.5\ R_\sun$ during CR 2231. A reverse color table is used to display images for enhancing the coronal structures. \label{fig:lasco_carrington}}
\end{center}
\end{figure*}

The Carrington map of coronagraph observations describes the coronal structures during the entire CR, presenting the evolution of the streamers, plumes, and CHs. It is composed of slices extracted at a specific heliocentric distance located in a series of coronagraphic images. The topological configurations of HCSs near the outer boundaries are model-dependent and sensitive to the selection of free parameters. The comparisons between HCSs and positions of streamers in Carrington maps can be used to evaluate the performance of different models, which can further be used to optimize model inputs, such as magnetograms, and free parameters \citep{Sasso2019}. Figure \ref{fig:lasco_carrington} shows the east (panel A) and west (panel B) limb Carrington maps derived from the LASCO C2 images at $2.5\ R_\sun$ during CR 2231. The date corresponding to Carrington's longitude is marked at the top of each panel. They are displayed using a reverse color table to enhance the visibility of coronal structures, where dark regions correspond to bright streamers.

\subsection{In-situ observations}

\begin{figure*}[htb!]
\begin{center}
\includegraphics[width=0.85\textwidth]{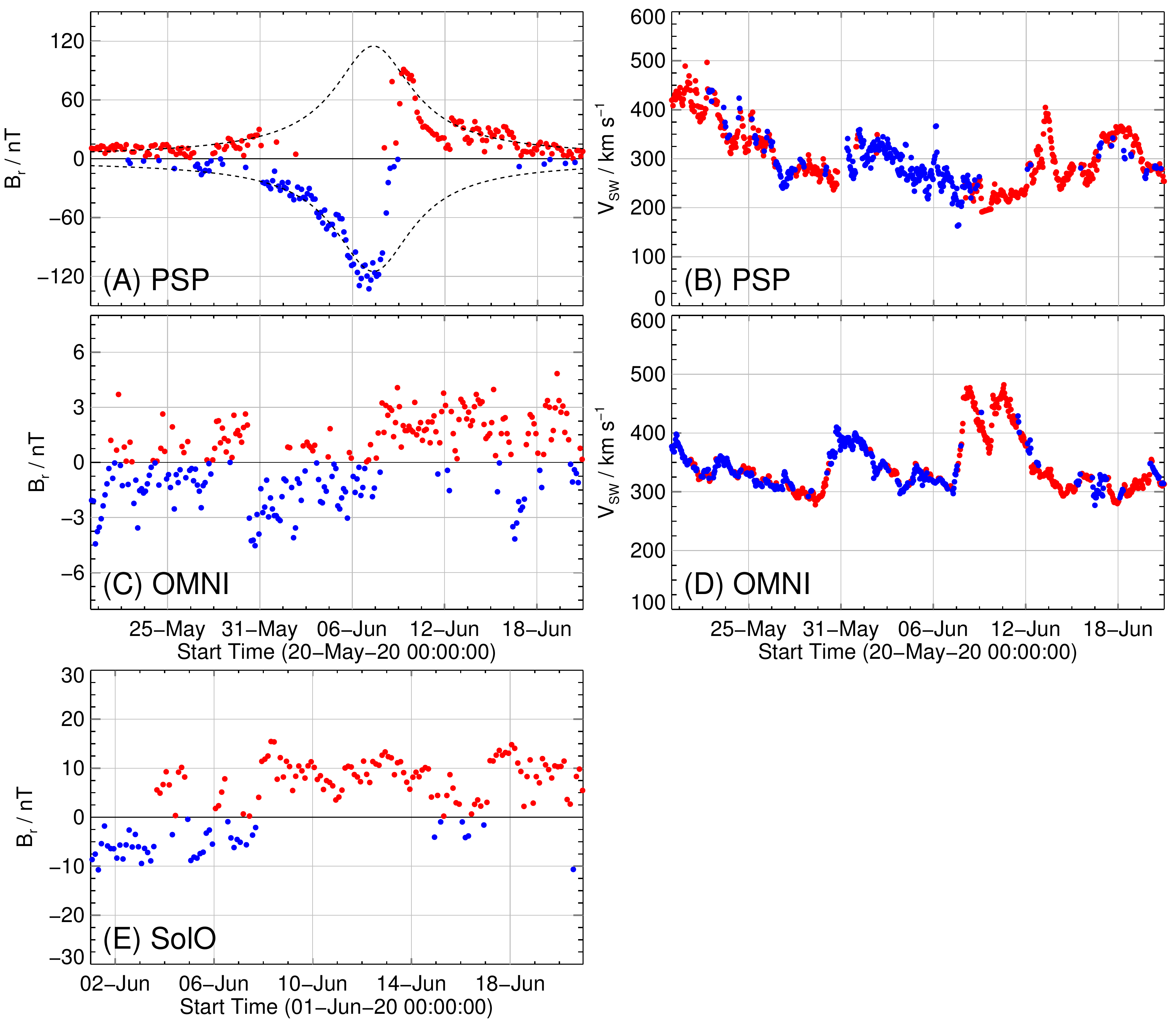}
\caption{IMFs and solar wind speeds in-situ measured by the PSP (A, B), the OMNI (C, D) and the SolO (E) during CR 2231. Three-hour averaging radial component $B_r$ of the IMFs derived from the FIELDS (A), OMNI (C), and MAG (E) for enhancing the polarity transition of IMFs. Proton bulk speed from the SWEAP (B) and OMNI (D) with a 1-hour cadence. Red and blue dots represent positive and negative polarities, respectively. Dashed lines in panel (A) represent the inverse square relationship between IMFs and the heliocentric distance of the PSP, $B_r \propto 1/r^2$. \label{fig:insitu_obs}}\end{center}
\end{figure*}

With the launch of PSP and SolO, plasma parameters can be measured in situ near the sun. Figure \ref{fig:insitu_obs} presents measurements of IMFs and solar wind speeds from the PSP (panels A, B), the OMNI datasets (panels C, D), and the SolO (panel E). Red dots in each panel indicate the polarities of the IMFs are positive ($B_r > 0$), and the direction of the IMFs is radially outward from the sun. Blue dots indicate the polarities are negative ($B_r < 0$).

There are four scientific payloads onboard the PSP \citep{Fox2016} launched on August 12, 2018, which is the first spacecraft to detect the sun with the closest distance. Measurements of the Electromagnetic Fields Investigation \citep[FIELDS,][]{Bale2016} and Solar Wind Electrons Alphas and Protons \citep[SWEAP,][]{Kasper2016} during the fifth solar encounter (E5) are used in this work. During this period, the PSP reached its perihelion of $27.9\ R_\sun$ on June 7. We pre-processed the original data to remove spikes and abnormal fluctuations, same as \citet{Badman2020}. Full cadence data are binned into 1 hour, and the modal values in each bin are represented as the observation values during this hour. To significantly observe the polarity transition of the magnetic fields, IMFs after pre-processing are averaged for three hours to smooth them. Figure \ref{fig:insitu_obs} (A) \& (B) show the radial component $B_r$ of the IMFs and the speed of proton bulk $V_{\rm SW}$, respectively, from May 20 to June 20, 2020. The dashed lines in panel (A) represent the relationship of the inverse square ratio between $B_r$ and the heliocentric distance $r$, $B_r \propto 1/r^2$. The averaged solar wind speed $V_{\rm SW}$ measured by the PSP is $300\ {\rm km\ s^{-1}}$.

The OMNI datasets \citep{Papitashvili2014} integrate numerous plasma parameters near the Earth, such as magnetic field, speed, temperature, and provisional activity indices of the sun. The hourly magnetic field and bulk flow speed from May 20 to June 20, 2020, are used in this work, produced by averaging multi-spacecraft in-situ measurements. Figure \ref{fig:insitu_obs} (C) shows the radial component $B_r$ of the IMFs performed in a 3-hour averaging consistent with the PSP, and panel (D) shows the bulk flow speed $V_{\rm SW}$. The averaged magnetic field strength $|B_r|$ is $1.85\ {\rm nT}$, and the averaged solar wind speed $V_{\rm SW}$ is $342\ {\rm km\ s^{-1}}$. 

There are ten payloads onboard the SolO launched on February 10, 2020, which is the most complex spacecraft combining in-situ and remote-sensing observations. The SolO can achieve a minimum perihelion distance of $0.28\ {\rm AU}$. It can reach a maximum orbital inclination of $24\,\degr$ away from the ecliptic. The radial component $B_r$ of the IMFs, shown in Figure \ref{fig:insitu_obs} (E), is in-situ measured by the Magnetometer \citep[MAG,][]{Horbury2020} in normal mode from June 1 to June 20, 2020. During this period, the SolO operates on an orbit with an average heliocentric distance of $0.52\ {\rm AU}$, and an orbital inclination of $5.61\,\degr$. The same pre-processing with the PSP bins the full cadence magnetic data of the MAG into 1 hour and performs in a 3-hour averaging. The averaged value of $|B_r|$ is $7.80\ {\rm nT}$ during this period.

\section{Magnetic field models} \label{sec:mag_mod}

\subsection{Coronal magnetic field models} 

Magnetohydrostatics (MHS) equations are obtained from MHD equations ignoring time terms, describing the balance of the Lorentz force with the plasma pressure and the gravitational force. MHS equations are composed of a magnetostatic balance equation and Maxwell equations \citep{Low1985},
\begin{eqnarray}
\frac{1}{\mu_0}(\nabla \times \bm{B}) \times \bm{B} - \nabla p - \rho \nabla \Phi_{\rm g} &=& 0, \label{equ:mhs} \\
\nabla \cdot \bm{B} &=& 0, \label{equ:max1} \\
\nabla \times \bm{B} &=& \mu_0 \bm{J}, \label{equ:max2}
\end{eqnarray}
where $\mu_0$, $\bm{B}$, $p$, $\rho$, and $\bm{J}$ are the permeability of vacuum, the magnetic field vector, the pressure, the plasma density, and the electric current vector, respectively. $\Phi_{\rm g}$ is expressed as the gravitational potential.

\subsubsection{Potential field source surface}

The PFSS model assumes that the electric current $\bm{J}$ is free in the corona, simultaneously ignoring the pressure and density terms in equation (\ref{equ:mhs}). Thus, the magnetic field $\bm{B}$ can be expressed by a magnetic scalar potential $\Phi_{\rm B}$, such that $\bm{B} = -\nabla \Phi_{\rm B}$, between the photosphere $R_\sun$ and a chosen radius $R_{\rm ss}$, called source surface. In the region $r > R_{\rm ss}$, the magnetic field lines gradually orient to the radial direction. According to Maxwell equations (\ref{equ:max1}) and (\ref{equ:max2}), $\Phi_{\rm B}$ satisfies the Laplace equation,
\begin{equation}
\nabla^2 \Phi_{\rm B} = 0. \label{equ:lap}
\end{equation}

The analytical solution of the PFSS model can be obtained by solving the Laplace equation (\ref{equ:lap}) in a spherical coordinate. Three components of the magnetic field ($B_r$, $B_\theta$, $B_\phi$) at a location ($r$, $\theta$, $\phi$) can be given by \citep{Freeland1998, Li2021},
\begin{eqnarray}
B_r &=& \sum_{n=1}^{N_{\rm max}}\left[(n + 1)\mathcal{A}+n\mathcal{B}\cdot\mathcal{C}\right] \times \nonumber \\
& & \sum_{m=0}^{n}P_n^m(g_{nm}\cos m\phi+h_{nm}\sin m\phi),  \\
B_\theta &=& -\sum_{n=1}^{N_{\rm max}}\left[\mathcal{A}-\mathcal{B}\cdot\mathcal{C}\right] \times \nonumber \\
& & \sum_{m=0}^{n}\frac{dP_n^m}{d\theta}(g_{nm}\cos m\phi+h_{nm}\sin m\phi), \\
B_\phi &=& \sum_{n=1}^{N_{\rm max}}\left[\mathcal{A}-\mathcal{B}\cdot\mathcal{C}\right] \times \nonumber \\
& & \sum_{m=0}^{n}P_n^m\frac{m}{\sin \theta}(g_{nm}\sin m\phi-h_{nm}\cos m\phi), 
\end{eqnarray}
where the functions $\mathcal{A}$, $\mathcal{B}$, and $\mathcal{C}$ in terms of heliocentric distance $r$ can be expressed by,
\begin{equation}
\mathcal{A} = \left(\frac{R_\sun}{r}\right)^{n+2}, \mathcal{B} = \left(\frac{R_\sun}{R_{\rm ss}}\right)^{n+2}, \mathcal{C} = \left(\frac{r}{R_{\rm ss}}\right)^{n-1}.
\end{equation}
$P_n^m$ is a function of $\cos \theta$ represented the Legendre polynomials, $g_{nm}$ and $h_{nm}$ are spherical harmonic coefficients obtained by boundary conditions, and $N_{\rm max}$ is the maximum order of $g_{nm}$ and $h_{nm}$. The calculation accuracy of the CMFs depends on $N_{\rm max}$, while results may appear as non-physical structures due to Legendre polynomials and trigonometric functions \citep{Toth2011ApJ}. To prevent the disadvantages of the analytical method and improve the computational efficiency, the finite-difference method \citep{Yeates2018,Stansby2020} is used to solve the Laplace equation (\ref{equ:lap}) in this work. 

\subsubsection{Potential field current sheet}

The PFCS model extends the CMFs from the source surface $R_{\rm ss}$, as derived by the PFSS model, to the higher corona $R_{\rm scs}$ by solving the Laplace equation (\ref{equ:lap}). To achieve this, the radial component $B_r$ of the magnetic field on the source surface $R_{\rm ss}$ serves as the lower boundary condition for the PFCS model. To construct structures of coronal current sheets, a polarity reversal method is applied in this model to induce the magnetic fields to point outward. Specifically, if $B_r < 0$, then $B_r$, $B_\theta$, and $B_\phi$ will be replaced by $-B_r$, $-B_\theta$, and $-B_\phi$, respectively. Once the magnetic fields in the region ($R_{\rm ss} < r \leq R_{\rm scs}$) are obtained, their original orientations will be restored. Instead of using the harmonic approach in this work, a finite-difference potential field solver \citep{Caplan2021ApJ} is applied to calculate the PFCS model.

The coupling of the PFSS and PFCS models can effectively reduce the difference between the equatorial and polar magnetic fields, reproducing the latitudinally independent variation of the radial component $B_r$ of the magnetic fields observed by Ulysses \citep{Smith1995}. It also improves the directions of coronal structures to match the observations \citep{Schatten1972}. The heights of the source surface $R_{\rm ss}$ in the PFSS model and the outer boundary $R_{\rm scs}$ in the PFCS model are crucial free parameters that significantly influence the extrapolation results. Further detailed discussions are presented in Section \ref{sec:results}.

\subsubsection{Current sheet source surface}

The CSSS model solves the same MHS equation (\ref{equ:mhs}-\ref{equ:max2}) as the PFSS model, with additional consideration of the horizontal current effect in the corona. The corona is divided into three regions, namely the inner, middle, and outer regions, based on the assumption of two surfaces: the cusp surface and the source surface. Three free parameters are contained in this model: the source surface height $R_{\rm ss}$, the cusp surface height $R_{\rm cs}$, and the current parameter $a$ describing the length scale of the horizontal current.

\citet{Bogdan1986} obtained the coronal electric current $\bm{J}$ and magnetic field $\bm{B}$ by solving the MHS equations (\ref{equ:mhs}-\ref{equ:max2}),
\begin{eqnarray}
\bm{J} &=& \frac{1}{\mu_0r} \left[1-\eta(r)\right] \left[\frac{1}{\sin\theta}\frac{\partial^2\Phi}{\partial \phi \partial r}\hat{\theta} - \frac{\partial^2\Phi}{\partial \phi \partial r}\hat{\phi}\right] \label{equ:current}, \\
\bm{B} &=& -\eta(r)\frac{\partial \Phi}{\partial r}\hat{r} - \frac{1}{r}\frac{\partial \Phi}{\partial \theta}\hat{\theta} - \frac{1}{r\sin \theta}\frac{\partial \Phi}{\partial \phi}\hat{\phi}, \label{equ:magnetic}
\end{eqnarray}
where $\eta(r)$ is a function of the current parameter $a$ and the heliocentric distance $r$, given by $\eta(r) = (1 + a/r)^2$. $\Phi$ is a potential-like function, and its values in different regions are determined by the boundary conditions. When the value of $a$ is $0$, it means that the assumed current in the corona is ignored.

In the inner region between the photosphere $R_{\sun}$ and cusp surface $R_{\rm cs}$, the height of the helmet streamer cusp \citep{Koutchmy1992} is usually higher than the pseudo-streamer \citep{Wang2007} observed in coronagraphic images, due to different current formation mechanisms. The height of the helmet streamer cusp can serve as a reference for selecting the parameter $R_{\rm cs}$. The function $\Phi$ in (\ref{equ:current}) and (\ref{equ:magnetic}) can be expressed as \citep{Zhao1994SoPh, Zhao1995, Koskela2019},
\begin{equation}
\Phi = \sum_{n=1}^{N_{\rm max}}\sum_{m=0}^{n} R_n P_n^m (g_{nm}\cos m\phi + h_{nm}\sin m\phi), \label{equ:inner}
\end{equation}
\begin{equation}
R_n = \frac{(1+a)^n}{(n+1)(r+a)^{n+1}},
\end{equation}
where the spherical harmonic coefficients $g_{nm}$ and $h_{nm}$ are calculated based on the photospheric synoptic maps.

In the middle region between the cusp surface $R_{\rm cs}$ and the source surface $R_{\rm ss}$, the CMF lines gradually orient towards the radial direction. The function $\Phi^c$ in this region can be expressed as \citep{Zhao1994SoPh, Zhao1995, Koskela2019},
\begin{equation}
\Phi^c = \sum_{n=1}^{N_{\rm max}^c}\sum_{m=0}^{n}R_n^c P_n^m (g_{nm}^c\cos m\phi + h_{nm}^c\sin m\phi), \label{equ:phi_c}
\end{equation}
\begin{eqnarray}
R^c_n &=& \left(\frac{n+1}{R_{\rm cs}^2 (R_{\rm cs}+a)^n} + \frac{n(R_{\rm cs}+a)^{n+1}}{R_{\rm cs}^2 (R_{\rm ss}+a)^{2n+1}} \right)^{-1} \times \nonumber  \\
& & \left(\frac{1}{(r+a)^{n+1}}-\frac{(r+a)^n}{(R_{\rm ss}+a)^{2n+1}} \right),
\end{eqnarray}
where the spherical harmonic coefficients $g_{nm}^c$ and $h_{nm}^c$ are determined by the radial component of the magnetic field on the cusp surface $R_{\rm cs}$ instead of the photosphere. The same polarity reversal technique used in the PFCS model is also applied to reverse the orientations of the CMFs wherever $B_r(R_{\rm cs}, \theta, \phi) < 0$ based on the inner boundary condition in the middle region.

In the outer region above the source surface $R_{\rm ss}$, magnetic field lines follow Parker spiral lines \citep{Parker1958}. The measurements of IMFs by near-Earth orbit spacecraft have provided observations that further confirm the accuracy of Parker spiral lines.

To sum up, once the functions $\Phi$ and $\Phi^c$ are obtained from equation (\ref{equ:inner}) and (\ref{equ:phi_c}), the CMFs can be calculated by equation (\ref{equ:magnetic}). Comparisons between the coupling model of PFSS+PFCS and the CSSS model are discussed in section \ref{sec:results}.

\subsection{Interplanetary magnetic field model} \label{subsec:parker}

Calculation of the IMF is obtained using the plasma flow parameters, assuming frozen-in fields and co-rotation with the sun. The magnetic field near the source surface satisfies,
\begin{equation}
B_\theta(R_{\rm ss}, \theta_{\rm ss}, \phi_{\rm ss}) = B_\phi(R_{\rm ss}, \theta_{\rm ss}, \phi_{\rm ss}) = 0.
\end{equation}
\citet{Parker1958} assumed that solar gravitation and outflow acceleration of plasma could be neglected above the source surface. Thus, the outflow speed of plasma is constant and can be represented by the solar wind speed $V_{\rm SW}$. In the interplanetary space, the co-rotation of plasma with the sun has a tangent speed $v_\phi$ besides a radial speed $v_r$,
\begin{eqnarray}
v_r(r, \theta, \phi) &=& V_{\rm SW}, \\
v_\theta(r, \theta, \phi) &=& 0, \\
v_\phi(r, \theta, \phi) &=& \Omega_\sun (r - R_{\rm ss}) \sin\theta,
\end{eqnarray}
where $\Omega_\sun$ is the rotational angular speed of the sun. Due to the magnetic freezing effect, both the plasma outflow and magnetic field lines present a spiral distribution. The corresponding relations between locations in the interplanetary space ($r$, $\phi$) and on the source surfaces ($R_{\rm ss}$, $\phi_0$) with the same latitude $\theta$ is,
\begin{equation}
\phi(r) = \phi_0 - \frac{\Omega_\sun}{V_{\rm SW}}(r - R_{\rm ss}).  \label{equ:parker_spiral}
\end{equation}
Plasma flows dominate the IMFs along spiral lines, so the strength of magnetic fields can be derived by,
\begin{eqnarray}
B_r(r, \theta, \phi) &=& B_r(R_{\rm ss}, \theta, \phi_0) \left(\frac{R_{\rm ss}}{r}\right)^2, \label{equ:brimf} \\
B_\theta(r, \theta, \phi) &=& 0, \label{equ:btimf} \\
B_\phi(r, \theta, \phi) &=& B_r(r, \theta, \phi) \left(\frac{\Omega_\sun}{V_{\rm SW}}\right)(r - R_{\rm ss})\sin \theta. \label{equ:bpimf}
\end{eqnarray}
It is worth noting that the three components of the IMF ($B_r$, $B_\theta$, $B_\phi$) derived from equations (\ref{equ:brimf})-(\ref{equ:bpimf}) do not strictly satisfy equation (\ref{equ:max1}).

\begin{figure}[htb!]
\begin{center}
\includegraphics[width=0.4\textwidth]{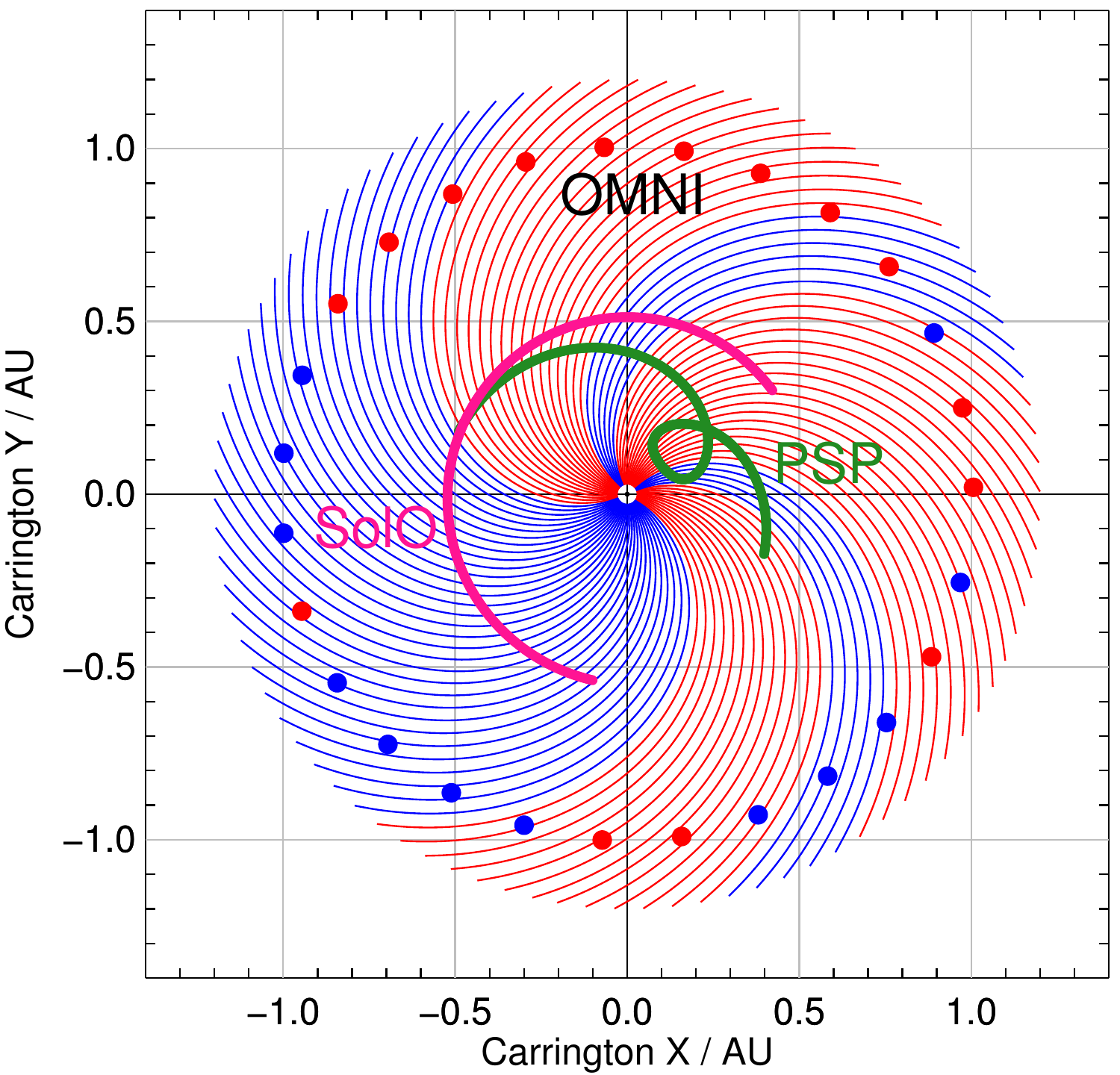}
\caption{Magnetic sectors in the Carrington coordinate, colored by magnetic polarities. Each field line is traced by the Parker spiral model and connected to the outer boundary $R_{\rm scs}=6.0\ R_\sun$ of the coupling model of PFSS+PFCS. Orbits of the PSP and SolO are represented by deep green and pink lines, respectively. Red and blue dots indicate the positive and negative polarities of the IMFs at $1.0\ {\rm AU}$, respectively, derived by a 24-hour average from OMNI datasets. \label{fig:parker}}
\end{center}
\end{figure}

Figure \ref{fig:parker} presents sectors of the IMFs in the Carrington coordinate colored with magnetic polarities, which are extrapolated from the outer boundary $R_{\rm scs}=6.0\ R_\sun$ derived by the PFSS+PFCS model to $1.2\ {\rm AU}$ using the HMI synoptic map. The polarities of each IMF line are traced by equation (\ref{equ:parker_spiral}) assuming a solar wind speed is $300\ {\rm km\ s^{-1}}$. Red lines indicate the positive polarity ($B_r > 0$), and the direction of the IMFs points outward of the sun. Blue lines indicate the negative polarity ($B_r < 0$), and the direction of the IMFs is towards the sun. Deep green and pink lines represent the orbital trajectories of the PSP and the SolO, respectively. The OMNI datasets combined merged hourly magnetic field is measured near the Earth, and averaged in a 24-hour cadence. Red and blue dots at $1.0\ {\rm AU}$ represent the positive and negative polarities of the IMFs, respectively.

\section{Results and comparisons} \label{sec:results}

\subsection{Comparisons among different free parameters}

The PFSS+PFCS model and the CSSS model have two and three free parameters, respectively, and the accuracy of the model results relies on the careful selection of these parameters. To make the total magnetic flux comparable, the source surface $R_{\rm ss}$ of the PFSS model is chosen to be consistent with the cusp surface $R_{\rm cs}$ of the CSSS model \citep{Koskela2019}. For comparison purposes, the outer boundary $R_{\rm scs}$ of the PFSS+PFCS model is set to the same height as the source surface $R_{\rm ss}$ of the CSSS model. From the helmet streamers observed in Figure \ref{fig:euv_wl}, the cusp heights $R_{\rm cs}$ of them are less than $4.0\ R_\sun$ within FoVs of the K-Cor and LASCO C2. Meanwhile, \citet{Koskela2019} suggests that the optimal $R_{\rm cs}$ value of the CSSS model lies within $2.0-4.0\ R_\sun$ for $a \leq 0.1$ and $2.5-4.5\ R_\sun$ for $a \approx 1.0$. Therefore, in this work, the range for parameter $R_{\rm cs}$ is limited to $2.0-5.0\ R_\sun$, and the current parameter $a$ is restricted to $0.0-1.0\ R_\sun$. To enhance computational efficiency, both $N_{\rm max}$ and $N_{\rm max}^c$ in equation (\ref{equ:inner}) and (\ref{equ:phi_c}) are set as 9. Tables \ref{tab:pfcs} and \ref{tab:csss} list the different cases of free parameters selected in the PFSS+PFCS and CSSS models, respectively. The HMI synoptic map and GONG daily synoptic maps provide the radial component $B_r$ of the photospheric magnetic field.

\begin{deluxetable}{ccc}
\tablecaption{Different parameters in the coupling model of PFSS+PFCS, include the source surface $R_{\rm ss}$ (second column) and the outer boundary $R_{\rm scs}$ (third column), are applied to extrapolate CMFs. \label{tab:pfcs}}
\setlength{\tabcolsep}{8.0 mm}{
\tablehead{\colhead{Cases} & \colhead{$R_{\rm ss}/R_\sun$} & \colhead{$R_{\rm scs}/R_\sun$}}
\decimalcolnumbers
\startdata
1 & 2.0 & 6.0  \\
2 & 2.5 & 6.0  \\
3 & 4.0 & 6.0 \\
4 & 2.0 & 8.0 \\
5 & 2.0 & 10.0 \\
\enddata}
\end{deluxetable}

\begin{deluxetable}{cccc}
\tablecaption{Different parameters in the CSSS model, include the current parameter $a$ (second column), the cusp surface $R_{\rm cs}$ (third column), and the source surface $R_{\rm ss}$ (fourth column), are applied to extrapolate CMFs. \label{tab:csss}}
\setlength{\tabcolsep}{6.0 mm}{
\tablehead{\colhead{Cases} & \colhead{$a/R_\sun$} & \colhead{$R_{\rm cs}/R_\sun$} & \colhead{$R_{\rm ss}/R_\sun$}}
\decimalcolnumbers
\startdata
1 & 0.0 & 2.0 & 6.0  \\
2 & 0.0 & 2.5 & 6.0  \\
3 & 0.0 & 4.0 & 6.0 \\
4 & 0.5 & 2.0 & 6.0 \\
5 & 0.5 & 2.0 & 8.0 \\
6 & 0.5 & 2.0 & 10.0 \\
7 & 1.0 & 2.0 & 6.0
\enddata}
\end{deluxetable}

\begin{figure*}[htb!]
\begin{center}
\includegraphics[width=0.9\textwidth]{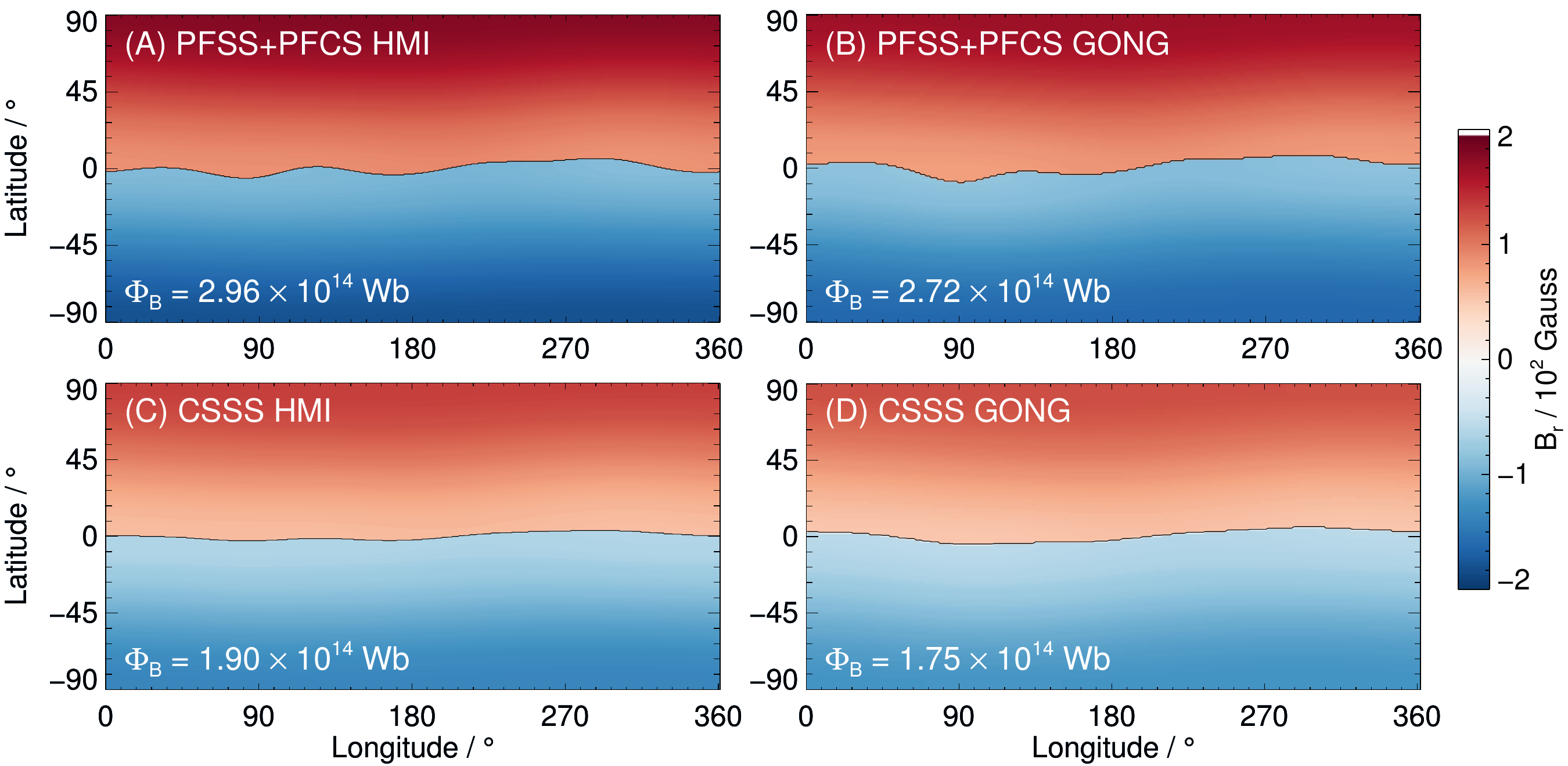}
\caption{Distributions of $B_r$ on the outer boundaries of models. (A) \& (B) are obtained by the PFSS+PFCS model with Case 2 ($R_{\rm ss} = 2.5\ R_\sun$, $R_{\rm scs} = 6.0\ R_\sun$) in Table \ref{tab:pfcs}. (C) \& (D) are obtained by the CSSS model with Case 2 ($R_{\rm cs} = 2.5\ R_\sun$, $R_{\rm ss} = 6.0\ R_\sun$, $a = 0.0$) in Table \ref{tab:csss}. The HMI synoptic map (left column) and GONG daily synoptic map (right column) serve as boundary conditions for the models. Black lines represent Heliospheric Current Sheets (HCSs). Total unsigned magnetic fluxes $\phi_{\rm B}$ are marked in each panel. \label{fig:pfss_csss_source}}
\end{center}
\end{figure*}

Figure \ref{fig:pfss_csss_source} shows distributions of $B_r$ near different outer boundaries obtained from the PFSS+PFCS model with Case 2 ($R_{\rm ss} = 2.5\ R_\sun$, $R_{\rm scs} = 6.0\ R_\sun$) in Table \ref{tab:pfcs} (panels A, B) and the CSSS model with Case 2 ($R_{\rm cs} = 2.5\ R_\sun$, $R_{\rm ss} = 6.0\ R_\sun$, $a = 0.0$) in Table \ref{tab:csss} (panels C, D), using the HMI synoptic map (left column) and the GONG daily synoptic map (right column) as boundary conditions. The total unsigned magnetic flux $\phi_{\rm B}$ near the source surface can be calculated by,
\begin{eqnarray}
\phi_{\rm B} = 4\pi R^2_{\rm ss} \langle \mid B_r(R_{\rm ss}) \mid \rangle.
\end{eqnarray}
From comparisons, $\phi_{\rm B}$ derived from the PFSS+PFCS model is $55\%$ larger than that from the CSSS model. Additionally, it is found that $\phi_{\rm B}$ obtained from the GONG is $8\%$ smaller than that from the HMI, for both models. Black lines represent HCSs on the outer boundaries, and the magnetic fields exhibit a large gradient in these regions. Meanwhile, we find that when increasing the distance between $R_{\rm ss}$ and $R_{\rm scs}$ for the PFSS+PFCS model, as well as between $R_{\rm cs}$ and $R_{\rm ss}$ for the CSSS model, the latitudinal profiles of the magnetic field strength $|B_r|$ on the outer boundaries tend to be a constant value (further comparisons are presented in Section \ref{sec:optimal}).

We further compare the total unsigned magnetic fluxes $\phi_{\rm B}$ and HCSs of the PFSS+PFCS and CSSS models with different cases. In the PFSS+PFCS model, when $R_{\rm scs}$ is a constant value of $6.0\ R_\sun$ and $R_{\rm ss}$ increases from $2.0\ R_\sun$ to $4.0\ R_\sun$, $\phi_{\rm B}$ decreases from $3.35 \times 10^{14}\ {\rm Wb}$ to $2.18 \times 10^{14}\ {\rm Wb}$ for the HMI synoptic map and from $3.08 \times 10^{14}\ {\rm Wb}$ to $2.01 \times 10^{14}\ {\rm Wb}$ for the GONG daily synoptic map. Additionally, the HCSs gradually flatten out. By only increasing $R_{\rm scs}$, the value of $\phi_{\rm B}$ decreases slightly. In the CSSS model, when either decreasing $R_{\rm cs}$ or increasing $a$, $\phi_{\rm B}$ increases, and the HCSs become more complex. However, when only increasing $R_{\rm ss}$, $\phi_{\rm B}$ remains constant, and the HCSs exhibit similar topological configurations. These results indicate that the total unsigned magnetic fluxes and HCSs of the CSSS model are mainly influenced by $R_{\rm cs}$ and $a$, instead of relying on $R_{\rm ss}$, which are consistent with \citet{Koskela2019}. 

\begin{figure*}[htb!]
\begin{center}
\includegraphics[height=0.32\textwidth]{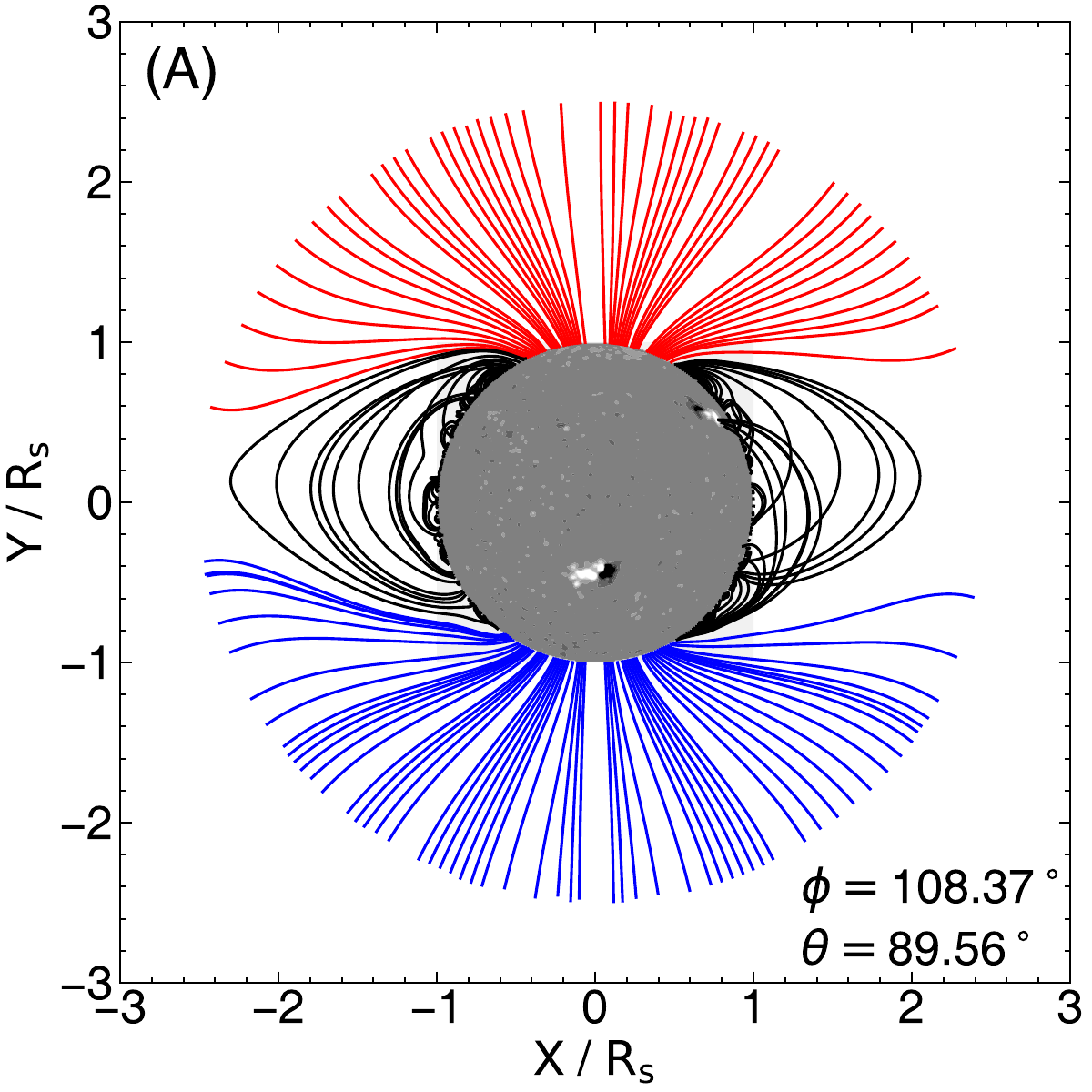}
\includegraphics[height=0.32\textwidth]{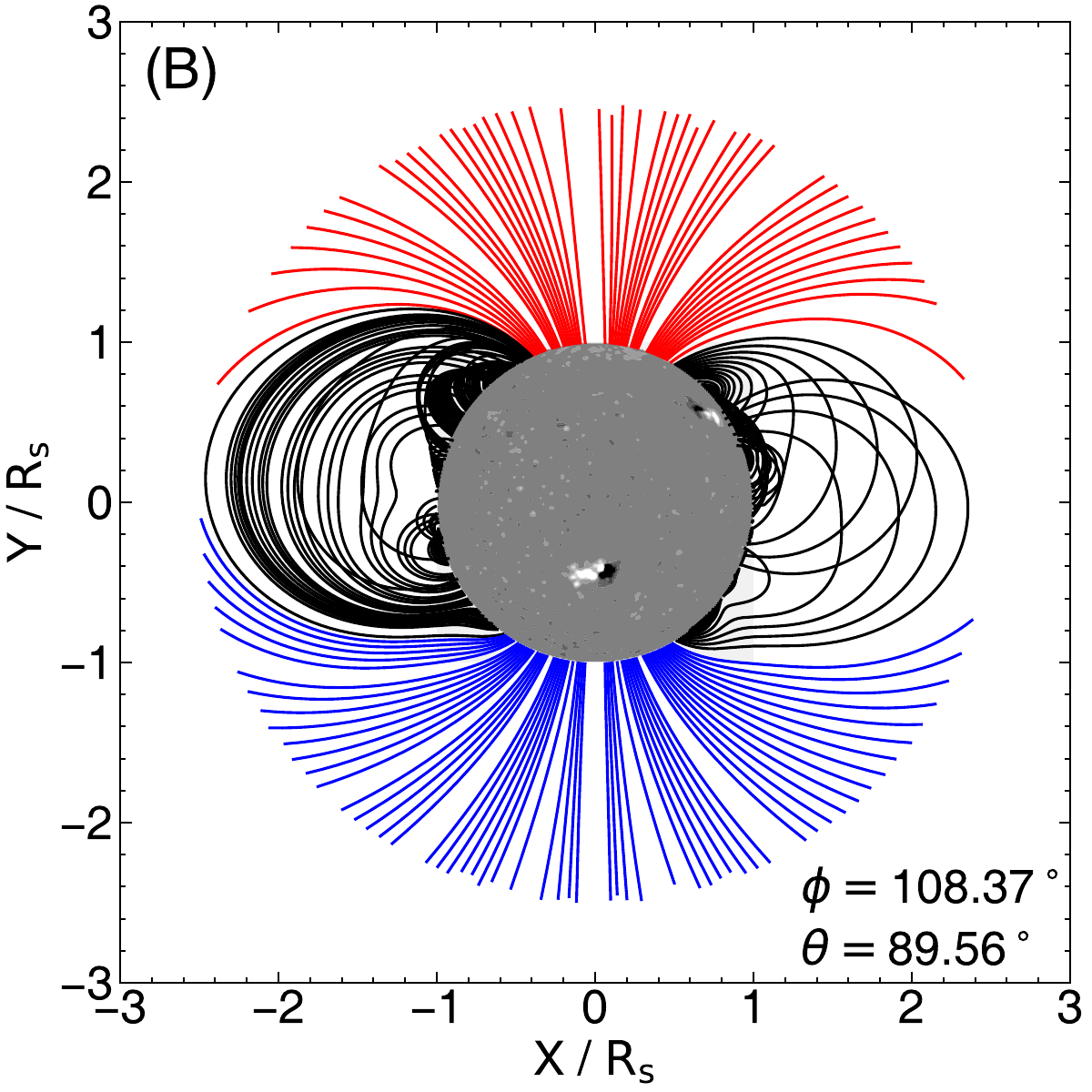} \\
\includegraphics[height=0.32\textwidth]{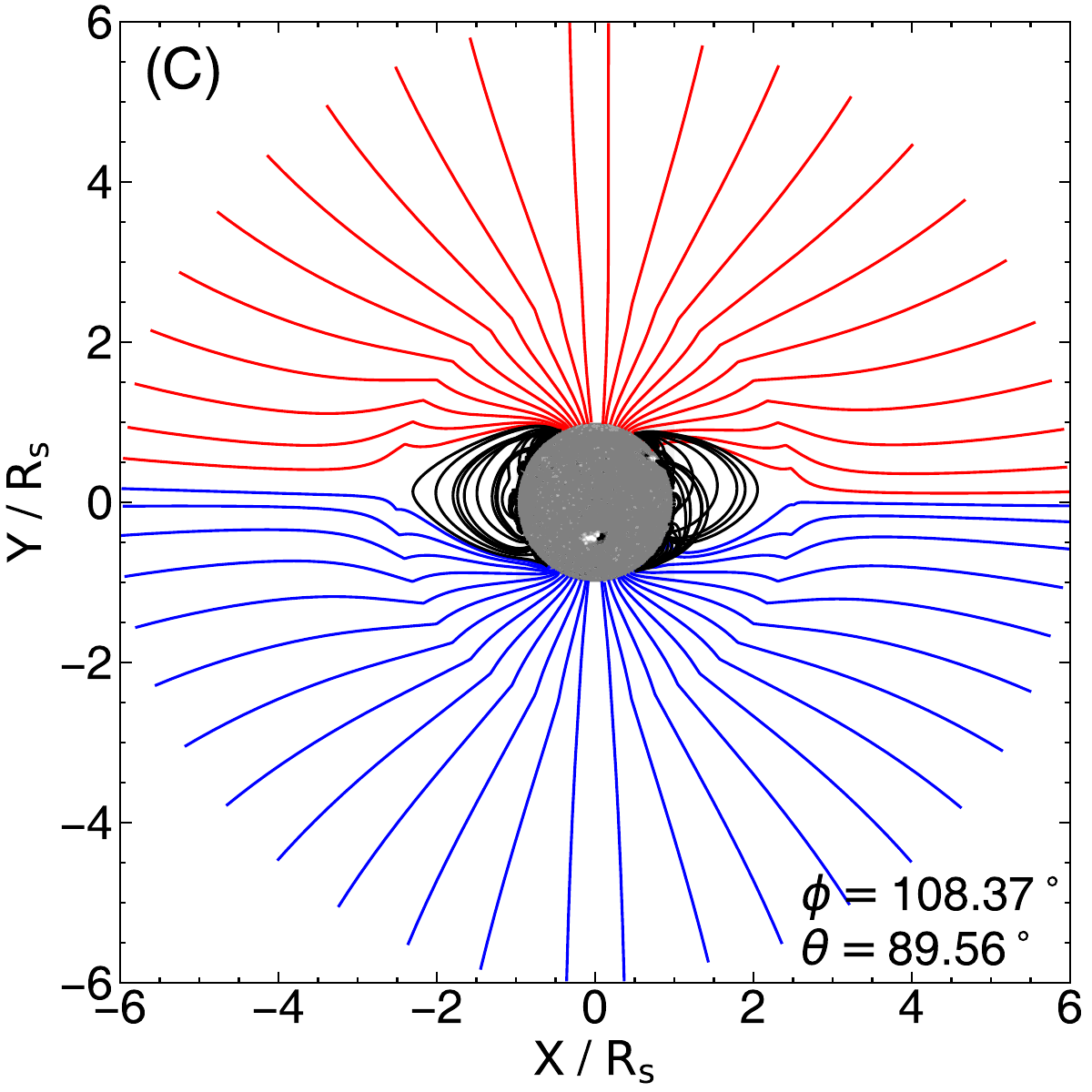}
\includegraphics[height=0.32\textwidth]{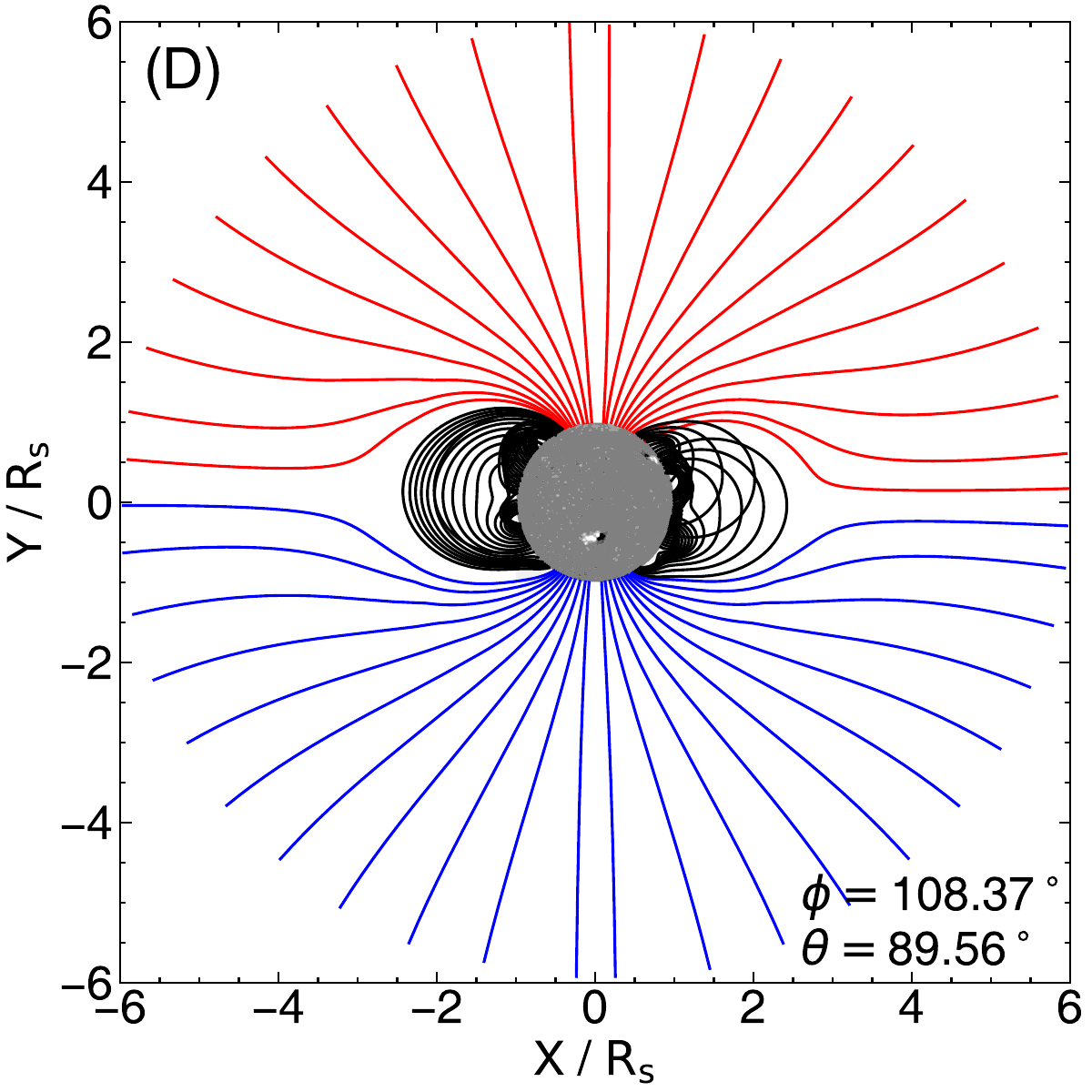}
\caption{CMF lines with different FoVs traced from the PFSS+PFCS model (A, C) with Case 2 in Table \ref{tab:pfcs} and the CSSS model (B, D) with Case 2 in Table \ref{tab:csss} are projected in the Plane-of-Sky (PoS) as viewed from Earth on June 9, 2020, using the GONG daily synoptic map. Positive and negative magnetic polarities are represented by red and blue lines, respectively, while black lines represent closed magnetic field lines. The longitude ($\phi$) and co-latitude ($\theta$) of the Earth's viewpoint in the Carrington coordinate are displayed in each panel. \label{fig:pfss_csss_mag_line}}
\end{center}
\end{figure*}

We also compare the CMF topological configurations obtained from the PFSS+PFCS and CSSS models on a nonuniform spherical mesh with dimensions of $61\times 181\times 361$ ($r$, $\theta$, $\phi$) in the FoV of $1.0-2.5\ R_\sun$ and $121\times 181\times 361$ ($r$, $\theta$, $\phi$) in the FoV of $1.0-6.0\ R_\sun$ using the GONG daily synoptic map. The free parameters for the two models are selected as Case 2 in Table \ref{tab:pfcs} and Case 2 in Table \ref{tab:csss}, respectively. Figure \ref{fig:pfss_csss_mag_line} shows the CMF lines with different FoVs traced from the PFSS+PFCS model (panels A, C) and the CSSS model (panels B, D) projected in the Plane-of-Sky (PoS), as viewed from the Earth on June 9, 2020. Red and blue lines represent the positive and negative magnetic polarities, respectively. CMF lines at middle and low latitudes, represented by black lines, form closed loops known as streamer arcades. They do not contribute to the total magnetic flux. The solar disk is replaced by the projected GONG map. In panel (B), the CMF lines of the CSSS model exhibit a higher degree of uniformity and smoothness compared to those in panel (A). The PFSS model can construct the cusp structure consistent with streamers. In panel (C), there is a discontinuity at the interface between the PFSS and PFCS models, which was also revealed by \citet{McGregor2008JGRA}. The reason for this phenomenon is that the PFSS model enforces the magnetic field to be entirely radial at the source surface (upper boundary of the PFSS), whereas CMFs calculated by the PFCS model exhibit a small non-radial component at the source surface (lower boundary of the PFCS). Conversely, the open magnetic field lines of the CSSS model shown in panel (D) appear notably smoother and more continuous than those in panel (C). Meanwhile, the locations of the horizontal current sheets present good consistency with each other.

Upon comparing the outer boundary distributions and the topological configurations of HCSs and magnetic field lines, we find that the CSSS model is essentially different from the PFSS+PFCS model. Although the current parameter $a$ is set to $0$, the CSSS model still maintains the original magnetic topology configurations. This discrepancy arises from the inclusion of helmet streamers and their associated current sheets in the CSSS model, while the PFSS+PFCS model solely assumes a potential field.

\subsection{Comparisons with remote-sensing observations}

\begin{figure*}[htb!]
\begin{center}
\includegraphics[width=0.75\textwidth]{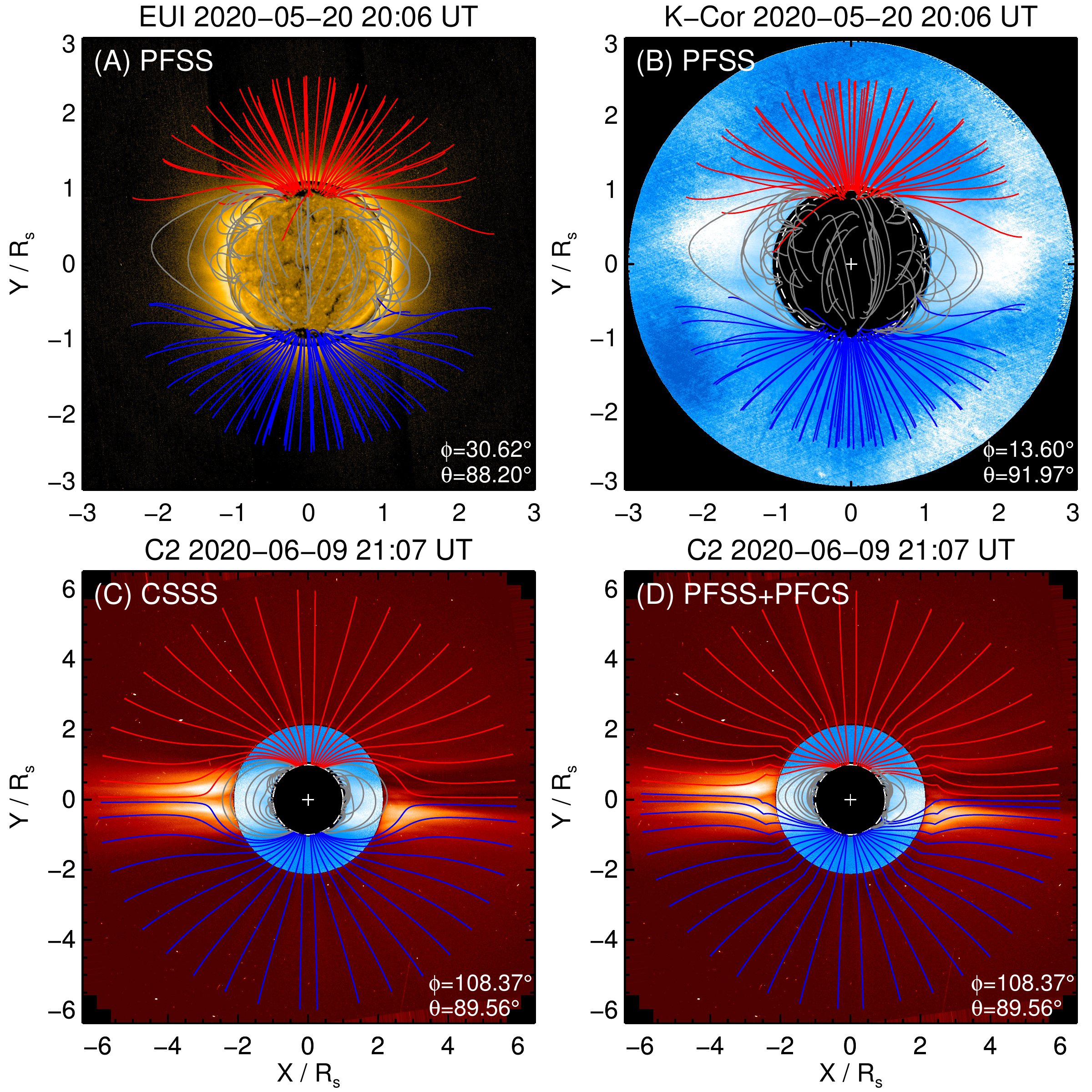}
\caption{Comparisons of CMF lines with EUI (A), K-Cor (B) and LASCO C2 (C, D) images. Red and blue lines are open magnetic field lines from positive and negative magnetic polarities, respectively, while gray lines represent closed magnetic field lines. The magnetic fields extrapolated by the PFSS+PFCS and CSSS models are traced to $6.0\ R_\sun$. Panels (A)-(C) use the HMI synoptic map, and panel (D) uses the GONG daily synoptic map. Viewpoints ($\phi$, $\theta$) of the SolO and the Earth are marked in each panel, respectively. \label{fig:euv_pfss_csss}}
\end{center}
\end{figure*}

To assess the model performance, we conduct a comprehensive evaluation by combining remote-sensing observations. Figure \ref{fig:euv_pfss_csss} shows the EUV and VL images observed by the EUI (panel A), K-Cor (panel B), as well as the composite image (panels C, D) of K-Cor and LASCO C2 on May 20 and June 9, 2020, overplotted with the magnetic field lines traced from the PFSS+PFCS and CSSS models using the HMI synoptic map (panels A, B, C) and GONG daily synoptic map (panel D), respectively. Red and blue lines represent open magnetic field lines rooted at positive and negative magnetic polarities, while gray lines represent closed magnetic field lines.

In Figure \ref{fig:euv_pfss_csss} (A) \& (B), the PFSS model utilizes free parameters with $R_{\rm ss} = 2.5\ R_\sun$, while the CSSS model (panel C) incorporates free parameters such as $R_{\rm cs} = 2.5\ R_\sun$, $R_{\rm ss} = 6.0\ R_\sun$, and $a = 0.2\ R_\sun$. Panel (D) presents the CMF lines extrapolated by the PFSS and PFCS models coupled in different regions. The PFSS model is limited to calculating CMFs in the low corona, whereas the PFCS and CSSS models can extend CMFs to heliocentric distances exceeding $10.0\ R_\sun$. In panel (A), the open magnetic lines, which are associated with plumes located at the north and south poles, connect to the dark EUV emission regions on the solar disk. This observation suggests that the primary source of the solar open magnetic field is the CHs where EUV emissions are relatively low. From comparisons shown in panels (B, C, D), the orientations of the CMF lines are in alignment with coronal rays, and the cusp heights of the closed CMFs closely resemble those of streamers. The results obtained from the two types of models using different synoptic maps are similar and effectively capture the true coronal structures.

\begin{figure*}[htb!]
\begin{center}
\includegraphics[width=0.9\textwidth]{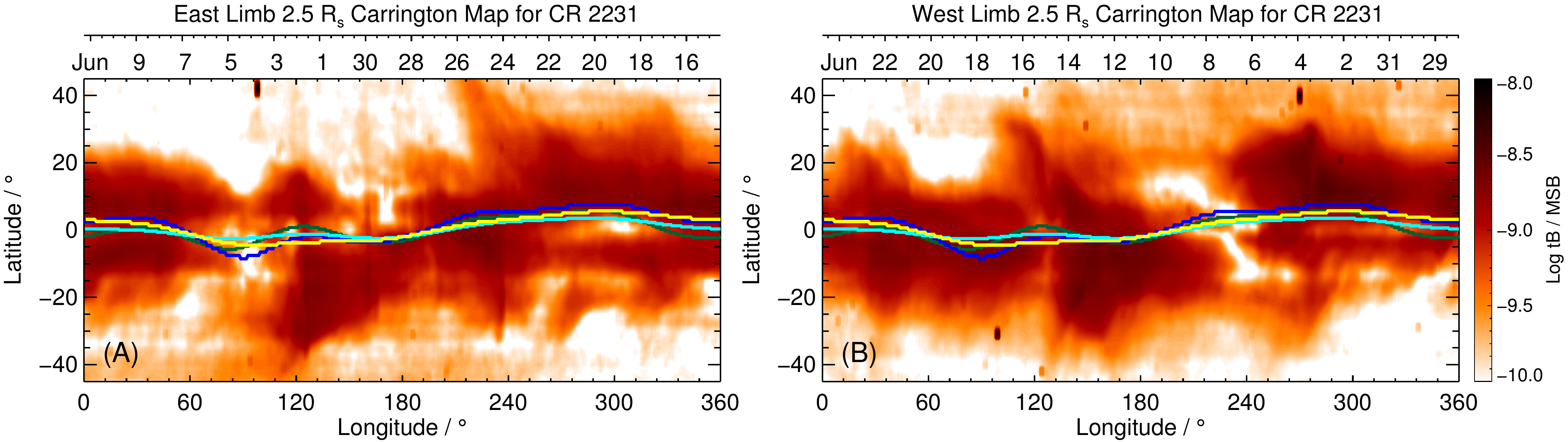} 
\caption{Carrington maps on the east (A) and west (B) limbs, generated by LASCO C2, overplotted HCSs near the outer boundaries of models shown in Figure \ref{fig:pfss_csss_source}. Green and blue lines represent calculations from the PFSS+PFCS model with Case 2 in Table \ref{tab:pfcs}, utilizing the HMI synoptic map and GONG daily synoptic maps, respectively. Cyan and yellow lines correspond to calculations made by the CSSS model with Case 2 in Table \ref{tab:csss}, utilizing the HMI and GONG maps, respectively. \label{fig:lasco_carr_csss_pfss}}
\end{center}
\end{figure*}

Figure \ref{fig:lasco_carr_csss_pfss} shows comparisons of the east (panel A) and west (panel B) limb Carrington maps with HCSs derived from models. The colored lines in each panel represent distinct HCSs that have been illustrated in Figure \ref{fig:pfss_csss_source}. We discovered that Carrington maps generated at varying heliocentric distances exhibit only variations in brightness, while the positions of the streamers remain constant. Therefore, we only show the results of Carrington maps at $2.5\ R_\sun$. The green and blue lines are computed using the PFSS+PFCS model, with the HMI synoptic map and GONG daily synoptic maps serving as their respective boundary conditions. Meanwhile, the cyan (derived from the HMI) and yellow (obtained from the GONG) lines are calculated using the CSSS model, which exhibits a comparably flatter profile in contrast to the PFSS+PFCS model. The evolved streamer structures are distributed on both sides of the HCSs.

\subsection{Comparisons with in-situ observations}

\begin{figure*}[htb!]
\begin{center}
\includegraphics[width=0.85\textwidth]{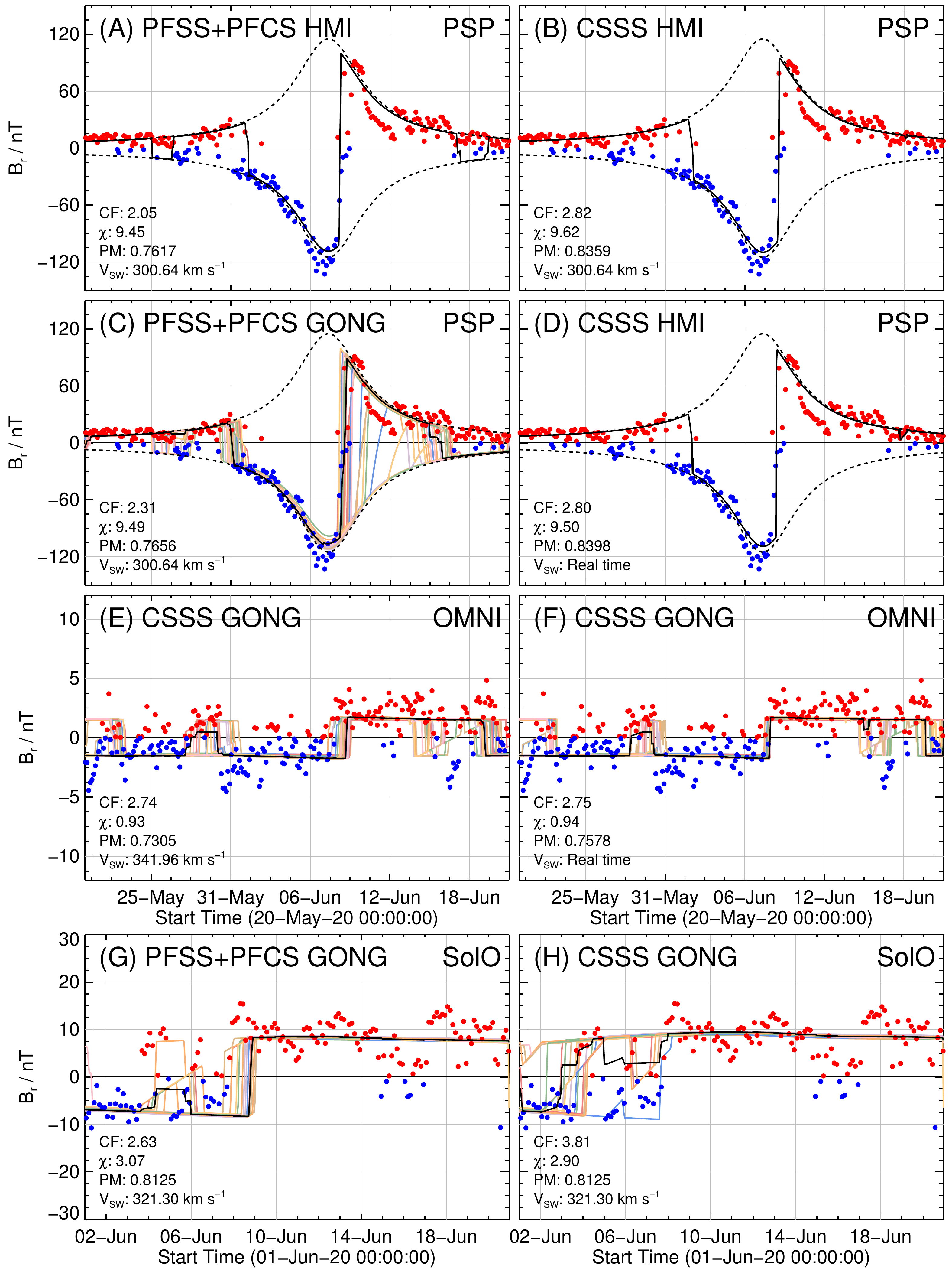}
\caption{Comparisons of the predicted IMFs (solid black lines) by the PFSS+PFCS and CSSS models with in-situ measurements (red and blue dots) obtained from the PSP (A)-(D), OMNI (E)-(F), and SolO (G)-(H). Colored lines are derived from the GONG daily synoptic maps, with a cadence of one day. Each panel includes the photospheric magnetograms used in models, three evaluation parameters, and the solar wind speed. The PFSS+PFCS model employs free parameters specified as Case 1 in Table \ref{tab:pfcs}, while the CSSS model uses Case 7 in Table \ref{tab:csss}. \label{fig:in_situ_comp}}
\end{center}
\end{figure*}

Numerous studies have consistently confirmed the underestimation of extrapolated IMFs with different models \citep{Wang1995, Linker2017, Riley2019a, Badman2021A&A}. However, the methods employed in these studies to estimate the IMFs primarily rely on the conservation of interplanetary magnetic flux. The averaged radial component of the IMF $B_{r}$ calculated by them at the heliocentric distance $r_{1}$ is,
\begin{eqnarray}
|B_{r}| &=& \frac{|\Phi_{\rm{open}}|}{4\pi r^2_{1}} \nonumber \\ 
&=& \frac{1}{4\pi}\left(\frac{R_{\rm{ub}}}{r_1}\right)^2 \int_{0}^{4\pi} |B_r(R_{\rm{ub}})|\ d\Omega, \label{equ:ofp}
\end{eqnarray}
where $R_{\rm{ub}}$ is the upper boundary of extrapolated models. When $r_{1}$ equals the heliocentric distance at SolO and near Earth, we estimate the ranges of $|B_r|$ to be $2.29-3.99\ {\rm nT}$ and $0.62-1.09\ {\rm nT}$, respectively, using the PFSS model with the value of $R_{\rm ss}$ in the range of $2.0 - 4.0\ R_\sun$. The predicted $|B_r|$ is significantly lower than the in-situ values of $7.80\ {\rm nT}$ measured by the SolO and $1.85\ {\rm nT}$ measured by the OMNI, respectively. Possible reasons for underestimation are summarized in \citet{Linker2017}: (1) the measured photosphere magnetic fields are systematically underestimated, particularly in the quiet and polar regions; (2) not all the open magnetic fields come from dark regions in EUV emission. In this work, we attempted to estimate the magnetic field by considering the variations of IMFs with latitude, longitude, and distance, using Parker spiral lines. Nevertheless, the results are still underestimated. Therefore, a Constant Factor (CF) is applied to scale the predicted IMFs to match the observed measurements. The smaller the CF value, the more accurate the calculated model results.

The degree of agreement in strength and polarity between the predicted IMFs (denoted as $O$) and the measured IMFs (denoted as $M$) is evaluated using the unsigned residual metric $\chi$ and Polarity Match (PM), respectively. $\chi$ is given by,
\begin{equation}
\chi(M, O) = \frac{1}{N}\sum_{i=1}^N\sqrt{(|M_i| - |O_i|)^2}.
\end{equation}
Both $M$ and $O$ are vectors of dimension $N$. The metric $\chi$ quantifies the level of quantitative consistency between models and observations regarding magnetic field strength. Moreover, the PM of IMFs serves as a crucial evaluation parameter for assessing the consistency in polarities. The smaller the value of $\chi$ and the larger the value of PM, the higher the consistency between the predicted and measured IMFs. The comparisons of the predicted IMFs derived from the PFSS+PFCS and CSSS models with in-situ measurements of the PSP, SolO, and OMNI are shown in Figure \ref{fig:in_situ_comp}. The free parameters of models are chosen as Case 1 listed in Table \ref{tab:pfcs} and Case 7 listed in Table \ref{tab:csss}, respectively. Note that these two cases are used for comparisons among different models and inputs. The derivation of the optimal parameters in PFSS+PFCS and CSSS models will be discussed in Section 4.4. The positive and negative polarities of the IMFs are represented by red and blue dots, respectively, while the solid black lines indicate the predicted IMFs from the models. Three evaluation parameters, CF, $\chi$, and PM, are indicated in each panel. The CF value is determined when the corresponding $\chi$ is optimal in each case.

In Figure \ref{fig:in_situ_comp}, the comparison between the radial component $B_r$ of the IMFs extrapolated from the outer boundary obtained by the PFSS+PFCS model and the in-situ measurements from the PSP is shown using the HMI synoptic map (panel A) and the GONG daily synoptic maps (panel C). In panel (C), colored lines are derived from the GONG daily synoptic maps with a cadence of one day, and the solid black lines are obtained by averaging the colored lines within $\pm 1.5$ days centered around the date of the relevant magnetograms corresponding to the observation date. The solar wind speed is assumed to be $V_{\rm SW} = 300\ {\rm km\ s^{-1}}$ derived by averaging the proton bulk speed measured in situ by the PSP during this period to generate Parker spiral lines connecting the PSP to the source surfaces. According to the CF and $\chi$, the predicted results from the HMI synoptic map are more consistent with the observations than the results from GONG daily synoptic maps. However, in terms of PM, GONG daily synoptic maps perform slightly better.

Subsequently, in Figure \ref{fig:in_situ_comp}, the results of the CSSS model are compared with the in-situ measurements from the SolO (panel H), and the OMNI (panel E) using GONG daily synoptic maps. The solar wind speed $V_{\rm SW}$ for the OMNI dataset is assumed to be $342\ {\rm km\ s^{-1}}$, which is obtained by averaging the hourly bulk flow speed. Since the SolO does not have particle speed data in this period, the solar wind speed is derived by averaging the proton bulk speed from both the PSP and the OMNI, resulting in $V_{\rm SW} = 321\ {\rm km\ s^{-1}}$. The CF at the location of the SolO is higher than that near Earth, despite using the same magnetograms, magnetic field models, and free parameters. Based on the same boundary conditions, the results of different models are also compared and presented in panels (A) and (B), as well as panels (G) and (H). For the PSP, the comparisons indicate that the IMF predictions from the PFSS+PFCS model outperform those from the CSSS model in terms of CF and $\chi$, while the CSSS model performs better in PM. Regarding the SolO, although $\chi$ obtained with the PFSS+PFCS model is larger than that with the CSSS model, the multiplied CF is smaller. This suggests that both models have advantages and disadvantages.

Then, the solar wind speed is substituted with in-situ measurements rather than an average value. Panels (D) and (F) of Figure \ref{fig:in_situ_comp} present the results calculated by the CSSS model using the real-time solar wind speed, shown in panels (B) and (D) of Figure \ref{fig:insitu_obs}, respectively. From the comparisons between panels (B) and (D), as well as panels (E) and (F), we find that incorporating real-time measurements of the solar wind speed can only improve the accuracy of extrapolation results slightly.

\begin{deluxetable*}{ccccccccccc}
\tablecaption{Three evaluation parameters, Constant Factor (CF), unsigned residual metric $\chi$, and Polarity Match (PM), obtained by comparing the predicted IMFs from the PFSS+PFCS model with different parameters listed in Table \ref{tab:pfcs}. The boundary conditions are HMI synoptic map and GONG daily synoptic maps, while the in-situ measurements are taken from the PSP, SolO, and OMNI. \label{tab:pfss_rss}}
\setlength{\tabcolsep}{3.0mm}{
\tablehead{\colhead{Cases} & \colhead{Magnetogram} & \multicolumn3c{CF} & \multicolumn3c{$\chi/{\rm nT}$} & \multicolumn3c{PM/\%} \\ 
\colhead{} & \colhead{} & \colhead{PSP} & \colhead{SolO} & \colhead{OMNI} & \colhead{PSP} & \colhead{SolO} & \colhead{OMNI} & \colhead{PSP} & \colhead{SolO} & \colhead{OMNI}}
\decimalcolnumbers
\startdata
1 & HMI & 2.05 & 2.39 & 1.81 & 9.45 & 2.95 & 0.91 & 76.17 & 81.25 & 61.72 \\
 & GONG & 2.31 & 2.63 & 1.99 & 9.49 & 3.07 & 0.98 & 76.56 & 81.25 & 69.14 \\ \cline{2-11} 
2 & HMI & 2.62 & 3.03 & 2.30 & 9.52 & 2.94 & 0.93 & 83.20 & 84.38 & 65.62 \\
 & GONG & 2.93 & 3.59 & 2.53 & 8.98 & 3.14 & 0.92 & 76.95 & 82.50 & 67.19 \\ \cline{2-11} 
3 & HMI & 6.20 & 6.73 & 5.40 & 9.95 & 2.87 & 0.92 & 75.78 & 71.25 & 66.80 \\
 & GONG & 6.85 & 7.65 & 5.99 & 11.47 & 3.14 & 0.94 & 73.44 & 71.25 & 71.09 \\ \cline{2-11} 
4 & HMI & 1.89 & 2.20 & 1.66 & 9.16 & 2.94 & 0.92 & 76.17 & 82.50 & 62.11 \\
 & GONG & 2.11 & 2.46 & 1.85 & 9.36 & 3.04 & 0.98 & 75.78 & 81.25 & 69.53 \\ \cline{2-11} 
5 & HMI & 1.81 & 2.13 & 1.59 & 9.28 & 2.94 & 0.91 & 76.95 & 82.50 & 61.72 \\
 & GONG & 2.03 & 2.38 & 1.78 & 9.27 & 3.07 & 0.97 & 75.78 & 81.88 & 68.36
\enddata
\tablecomments{The CF is determined when the corresponding $\chi$ is optimal.}}
\end{deluxetable*}
\begin{deluxetable*}{ccccccccccc}
\tablecaption{Three evaluation parameters, CF, $\chi$, and PM, obtained by comparing the predicted IMFs from the CSSS models with different parameters listed in Table \ref{tab:csss}. The boundary conditions are HMI and GONG maps, while the in-situ measurements are taken from the PSP, SolO, and OMNI. \label{tab:csss_case}}
\tablewidth{0pt}
\setlength{\tabcolsep}{3.0mm}{
\tablehead{\colhead{Cases} & \colhead{Magnetogram} & \multicolumn3c{CF} & \multicolumn3c{$\chi/{\rm nT}$} & \multicolumn3c{PM/\%} \\ 
\colhead{} & \colhead{} & \colhead{PSP} & \colhead{SolO} & \colhead{OMNI} & \colhead{PSP} & \colhead{SolO} & \colhead{OMNI} & \colhead{PSP} & \colhead{SolO} & \colhead{OMNI}}
\decimalcolnumbers
\startdata
1 & HMI & 2.82 & 3.29 & 2.51 & 9.89 & 2.91 & 0.91 & 78.91 & 71.25 & 67.19 \\
 & GONG & 3.19 & 3.77 & 2.78 & 10.01 & 3.13 & 0.91 & 76.95 & 75.63 & 69.53 \\ \cline{2-11}
2 & HMI & 3.85 & 4.39 & 3.37 & 9.87 & 2.90 & 0.90 & 77.73 & 71.25 & 68.36 \\
 & GONG & 4.27 & 4.92 & 3.74 & 10.42 & 3.06 & 0.91 & 75.39 & 71.25 & 71.09 \\ \cline{2-11}
3 & HMI & 13.13 & 13.79 & 11.56 & 9.70 & 2.81 & 0.89 & 77.34 & 71.25 & 68.75 \\
 & GONG & 14.71 & 14.89 & 12.66 & 11.22 & 2.86 & 0.91 & 73.44 & 71.25 & 71.88 \\ \cline{2-11}
4 & HMI & 2.78 & 3.19 & 2.42 & 9.06 & 3.05 & 0.92 & 81.64 & 75.00 & 67.97 \\
 & GONG & 3.09 & 3.81 & 2.70 & 9.45 & 3.23 & 0.91 & 76.95 & 80.00 & 68.75 \\ \cline{2-11}
5 & HMI & 2.41 & 2.79 & 2.12 & 9.53 & 2.92 & 0.90 & 81.25 & 71.25 & 68.75 \\
 & GONG & 2.70 & 3.20 & 2.34 & 9.82 & 3.23 & 0.91 & 76.56 & 77.50 & 69.14 \\ \cline{2-11}
6 & HMI & 2.27 & 2.63 & 2.00 & 9.80 & 2.92 & 0.90 & 81.25 & 71.25 & 68.75 \\
 & GONG & 2.53 & 2.97 & 2.23 & 10.03 & 3.04 & 0.91 & 76.56 & 71.25 & 69.92 \\ \cline{2-11}
7 & HMI & 2.82 & 3.17 & 2.43 & 9.62 & 2.90 & 0.91 & 83.59 & 81.88 & 65.23 \\
 & GONG & 3.13 & 3.81 & 2.74 & 9.21 & 2.90 & 0.93 & 77.34 & 81.25 & 73.05
\enddata}
\end{deluxetable*}

Furthermore, three evaluation parameters for all the cases in Tables 1 and 2 are listed in Table \ref{tab:pfss_rss} and \ref{tab:csss_case} for PFSS+PFCS and CSSS models, respectively. We find that the model performance is significantly influenced by the choice of free parameters. The established three evaluation parameters can assist us effectively in assessing the model performance and subsequently determining the optimal parameters.

\begin{enumerate}

\item The CF obtained near Earth is smaller than those at the positions of PSP and SolO in all cases for both models. It is worth noting that the CFs calculated from the HMI synoptic map are smaller than those obtained from the GONG daily synoptic maps, which may be attributed to the high accuracy measurements of the photospheric magnetic field by the HMI, resulting in a smaller underestimation of the IMFs. The IMF values predicted by the PFSS+PFCS model are larger than those predicted by the CSSS model. A possible reason is that the PFSS+PFCS model opens more magnetic flux, as shown in Figure \ref{fig:pfss_csss_mag_line}. The CF increases as the source surface $R_{\rm ss}$ increases or the outer boundary $R_{\rm scs}$ decreases for the PFSS+PFCS model. For the CSSS model, increasing the cusp surface $R_{\rm cs}$ or decreasing the source surface $R_{\rm ss}$ result in an increase in the CF. The current parameter $a$ has little effect on the CF. These results indicate that for the PFSS+PFCS model, $R_{\rm ss}$ should not exceed $4.0\ R_\sun$, and a larger value of $R_{\rm scs}$ should be selected. For the CSSS model, it is advisable to choose a smaller value for $R_{\rm cs}$ that still meets the observations, while opting for a larger value for $R_{\rm ss}$.

\item According to the comparison results of the PSP, it can be found that in most cases for both the PFSS+PFCS and CSSS models, the values of $\chi$ calculated from the HMI are smaller than those calculated from the GONG, while the values of PM are larger than those calculated from the GONG. The results suggest that the IMFs based on the HMI are more consistent with the measurements in terms of strength and polarity. By examining the three evaluation parameters in Table \ref{tab:pfss_rss} and \ref{tab:csss_case}, it is evident that the results of the PFSS+PFCS model have an advantage in predicting the strength of IMFs, as it can produce smaller values of CF and $\chi$. The CSSS model presents only a very slight advantage in predicting the polarities of IMFs, as its averaged PM value is larger than that of the PFSS+PFCS model by no more than $5.0\%$.

\item According to the comparison results of the SolO, the $\chi$ derived from the HMI is smaller than that from the GONG. In the PFSS+PFCS model, the averaged PM value obtained from the HMI is $0.9\%$ larger than that obtained from the GONG. However, in the CSSS model, the situation is the opposite, with the averaged PM value from the HMI being $2.8\%$ smaller than that from the GONG. This implies that the HMI provides slightly better results compared to the GONG for the PFSS+PFCS model, while the polarities of the IMFs are almost negligibly less accurate compared to those derived from the GONG for the CSSS model.

\item According to the comparison results of the OMNI, the GONG outperforms the HMI in terms of polarities for both models, as indicated by the larger PM values, but the strength predictions of the IMFs are slightly worse than those of the HMI. The possible reason is that although the GONG daily synoptic maps can reflect daily magnetic activities on the photosphere within one Carrington rotation, the spatial resolution and sensitivity is worse than that of the HMI.

\end{enumerate}

\subsection{Optimal free parameters of models} \label{sec:optimal}

To further investigate the optimal free parameters for the PFSS+PFCS and CSSS models, we use the in-situ measurements by PSP, SolO, OMNI, and Ulysses. We perform detailed calculations of the evaluation parameters based on the HMI map. The reason for choosing the HMI map is that the results obtained from the HMI and GONG are almost similar, and the required calculation time is short.

\begin{figure*}[htb!]
\begin{center}
\includegraphics[width=0.9\textwidth]{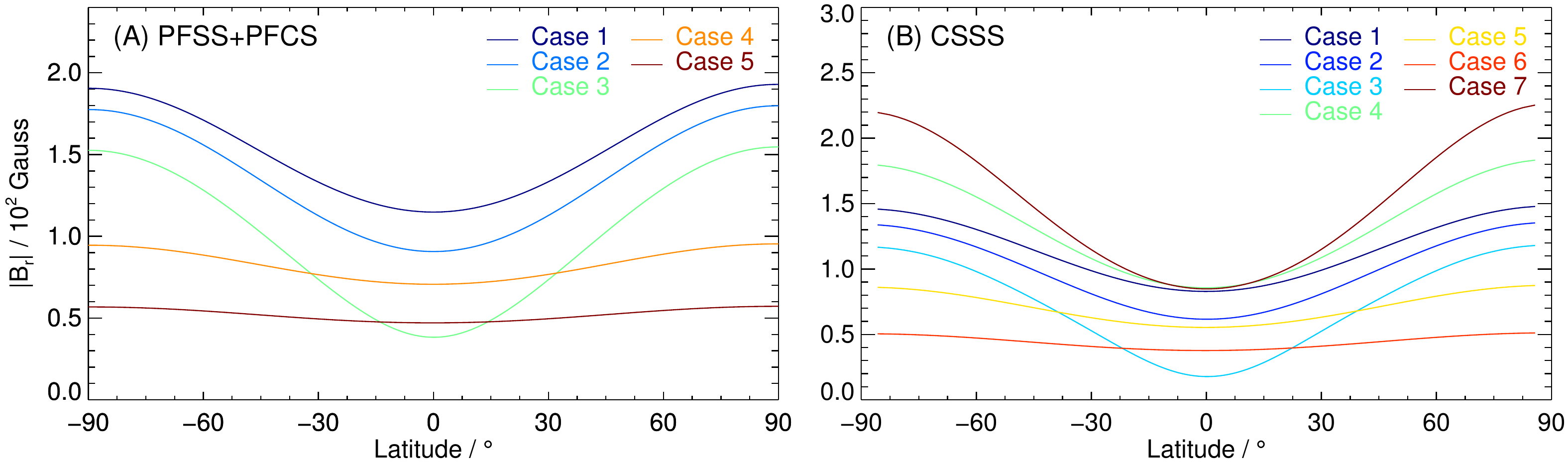}
\caption{Longitudinal averaged profiles of the unsigned $|B_r|$ on the outer boundaries of the PFSS+PFCS (A) and CSSS (B) models for all cases listed in Table \ref{tab:pfcs} \& \ref{tab:csss}, respectively, using the HMI synoptic map as boundary conditions. \label{fig:figure7}}
\end{center}
\end{figure*}

To ensure that the latitudinal variation of the magnetic field is consistent with Ulysses observations, we compare profiles averaging along the longitude of the outer boundaries under different parameter cases listed in Table \ref{tab:pfcs} and \ref{tab:csss}. Figure \ref{fig:figure7} presents profiles of the unsigned radial field $|B_r|$ as a function of latitude, which are obtained by the PFSS+PFCS model (panel A) and the CSSS model (panel B), respectively. For the PFSS+PFCS model, only increasing $R_{\rm ss}$ leads to an increase in the difference between the polar and equatorial magnetic fields, while only increasing $R_{\rm scs}$ results in almost a flat profile of $|B_r|$. When $a$ is a constant value, the profile variations of the CSSS model are similar to those of the PFSS+PFCS model. Decreasing $R_{\rm cs}$ or increasing $R_{\rm ss}$ can gradually smooth the $|B_r|$ variation in latitude. From the profiles of Cases 1, 4, and 7 shown in panel (B), it is worth noting that the value of $a$ influences the magnetic field strength at higher latitudes and has little effect on the field near the equator. When $R_{\rm scs}-R_{\rm ss}$ in the PFSS+PFCS model, or $R_{\rm ss}-R_{\rm cs}$ in the CSSS model, is larger than $8.0\ R_\sun$, a higher consistency can exist between the extrapolated magnetic fields and the Ulysses observations.

To further constrain the optimal parameters, in the PFSS+PFCS model, the source surface $R_{\rm ss}$ varies from $2.0$ to $5.0\ R_\sun$, with a step of $0.25\ R_\sun$, and the outer boundary $R_{\rm scs}$ is within the range of $3.5 - 15.0\ R_\sun$, with a step of $0.25\ R_\sun$. In the CSSS model, the cusp surface $R_{\rm cs}$ ranges from $2.0$ to $5.0\ R_\sun$, with a step size of $0.25\ R_\sun$. The source surface $R_{\rm ss}$ is selected within the range of $3.5$ to $15.0\ R_\sun$, with the same step size of $0.25\ R_\sun$. The current parameter $a$ is chosen between $0.0$ and $1.0\ R_\sun$, with a step of $0.5\ R_\sun$. Additionally, the relationship between $R_{\rm scs}$ and $R_{\rm ss}$ for the PFSS+PFCS model, as well as between $R_{\rm ss}$ and $R_{\rm cs}$ for the CSSS model, is constrained by $R_{\rm scs} \geq R_{\rm ss} + 1.5$ and $R_{\rm ss} \geq R_{\rm cs} + 1.5$, respectively.

\begin{figure*}[htb!]
\begin{center}
\includegraphics[width=1.0\textwidth]{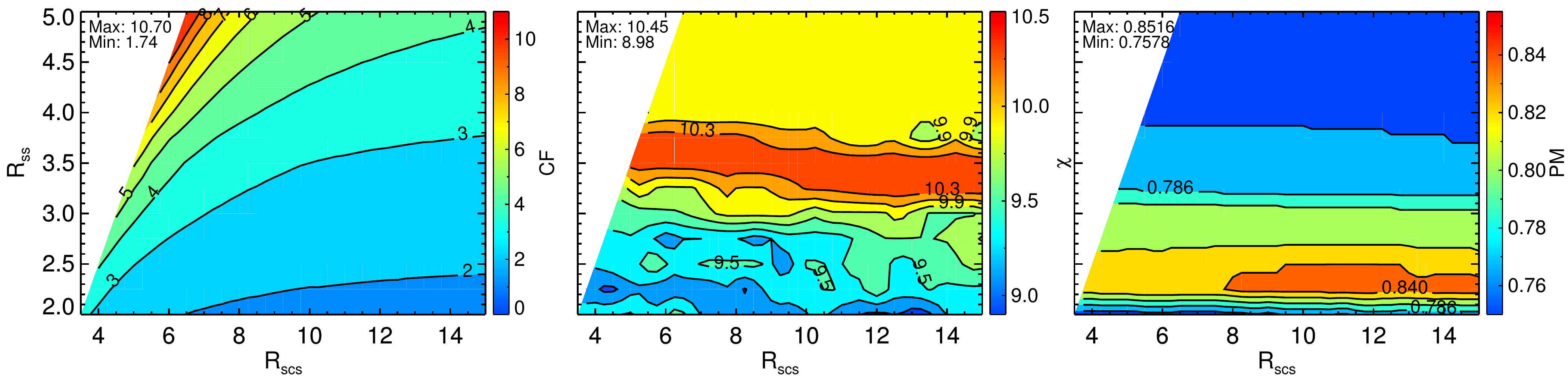}
\caption{Contour maps of the Constant Factor (CF, left column), the unsigned residual metric $\chi$ (middle column), and the Polarity Match (PM, right column) are derived by comparing the predicted IMFs of the PFSS+PFCS model utilizing the HMI synoptic map with the measurements obtained from the PSP. The evaluation parameters are plotted as functions of the source surface $R_{\rm ss}$ (vertical axis) and outer boundary $R_{\rm scs}$ (horizontal axis). Both the maximum and minimum values of each evaluation parameter are marked in the corners of each panel. \label{fig:psp_factor_pm_hmi_optimal}}
\end{center}
\end{figure*}

The distributions of three evaluation parameters, CF (left column), $\chi$ (middle column), and PM (right column), vary as a function of $R_{\rm ss}$ and $R_{\rm scs}$ obtained from the comparisons between the predicted IMFs with the PFSS+PFCS model using the HMI synoptic map and the in-situ measurements by the PSP, as shown in Figure \ref{fig:psp_factor_pm_hmi_optimal}. The maximum and minimum values are marked in each panel. It is evident that the CF monotonically increases with an increase in $R_{\rm ss}$ or a decrease in $R_{\rm scs}$. The minimum value of $\chi$ falls within the region with $R_{\rm ss}$ of $2.0 - 2.8\ R_\sun$ and $R_{\rm scs}$ of $4.0 - 14.0\ R_\sun$. Additionally, the maximum value of PM is $85.16\%$ in the region of $R_{\rm ss} = 2.2 - 2.5\ R_\sun$ and $R_{\rm scs} = 8.0 - 15.0\ R_\sun$. Considering the latitudinal-independent constraint of $|B_r|$, we suggest that the optimal range for the PFSS+PFCS model is $R_{\rm ss} = 2.2 - 2.5\ R_\sun$ and $R_{\rm scs} = 10.5 - 14.0\ R_\sun$.

\begin{figure*}[htb!]
\begin{center}
\includegraphics[width=1.0\textwidth]{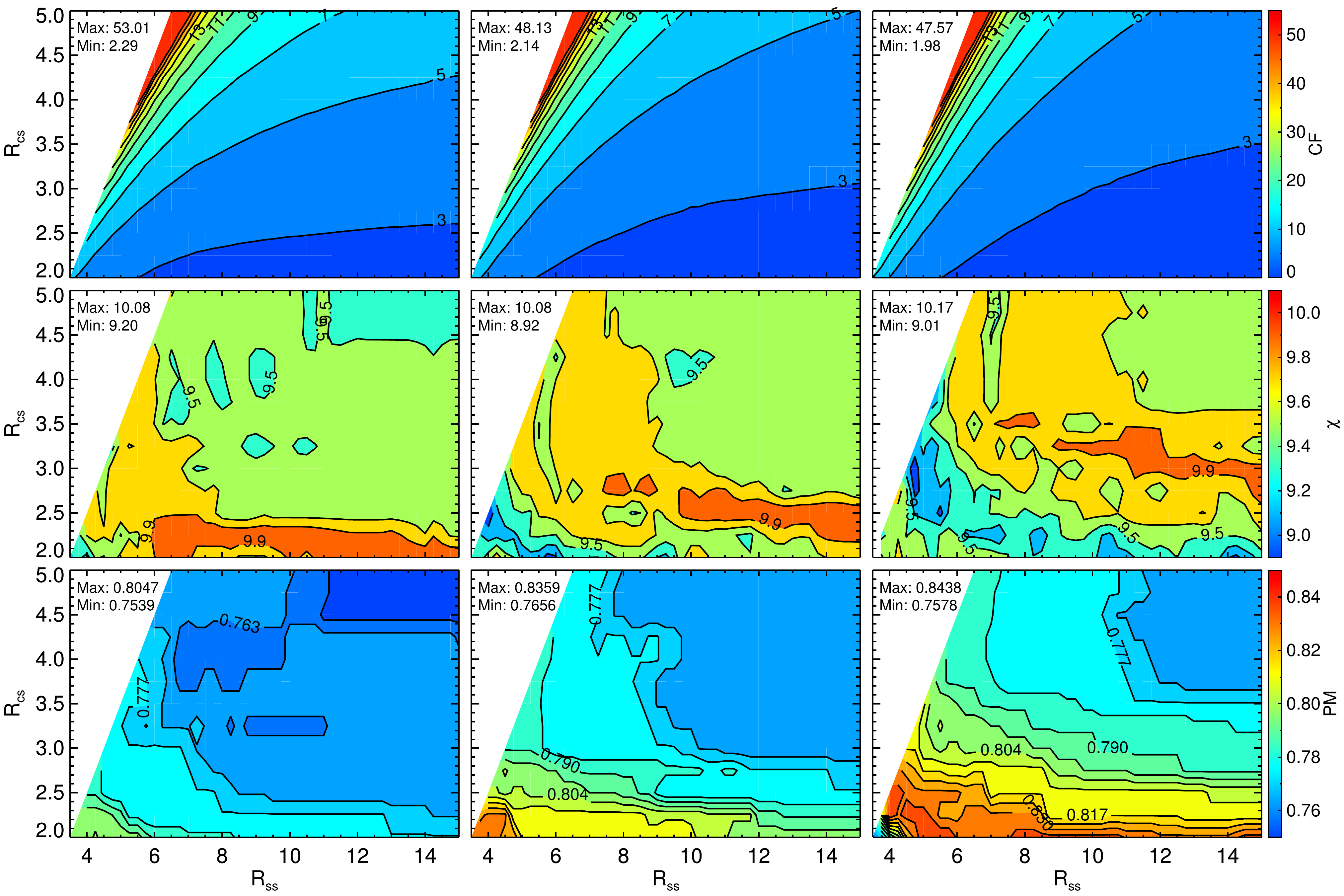}
\caption{Contour maps of the CF (top row), $\chi$ (middle row), and PM (bottom row) obtained by comparing the predicted IMFs from the CSSS model with the observations by PSP. Horizontal and vertical axes in each panel represent the source surface $R_{\rm ss}$ and cusp surface $R_{\rm cs}$, respectively. Each column has a different current parameter $a$, which are $0$ (left), $0.5\ R_\sun$ (middle), and $1.0\ R_\sun$ (right). Both the maximum and minimum values are marked in each panel. \label{fig:psp_factor_pm}}
\end{center}
\end{figure*}

Figure \ref{fig:psp_factor_pm} shows the evaluation parameters of the CSSS model using the HMI synoptic map, derived by comparing the calculated IMFs with the measurements of the PSP. Each column in Figure \ref{fig:psp_factor_pm} represents the results obtained with different current parameters $a$, namely $0$ (left column), $0.5\ R_\sun$ (middle column), and $1.0\ R_\sun$ (right column), respectively. The top, middle, and bottom rows represent the contour maps of the CF, $\chi$, and PM, respectively. The maximum and minimum values in each panel are indicated. The distributions of the three evaluation parameters are similar for different values of $a$. For the CF, when $R_{\rm ss}$ is relatively small, increasing $R_{\rm cs}$ leads to a steeper increase in the CF compared to when $R_{\rm ss}$ is large. In a region characterized by smaller $R_{\rm cs}$ and larger $R_{\rm ss}$, the CF has lower values. In terms of $\chi$, the optimal values are distributed in the region where $R_{\rm cs}$ is relatively small, and this region expands with increasing $a$. An optimal value of $a$ is $1.0\ R_\sun$. The distribution of PM follows a pattern opposite to that of $\chi$. Together with the constraint of the Ulysses observations, we have determined that the optimal range of free parameters for the CSSS model is $R_{\rm cs} = 2.0 - 2.4\ R_\sun$ and $R_{\rm ss} = 10.4 - 14.7\ R_\sun$ when $a = 1.0\ R_\sun$.

\begin{figure*}[htb!]
\begin{center}
\includegraphics[width=1.0\textwidth]{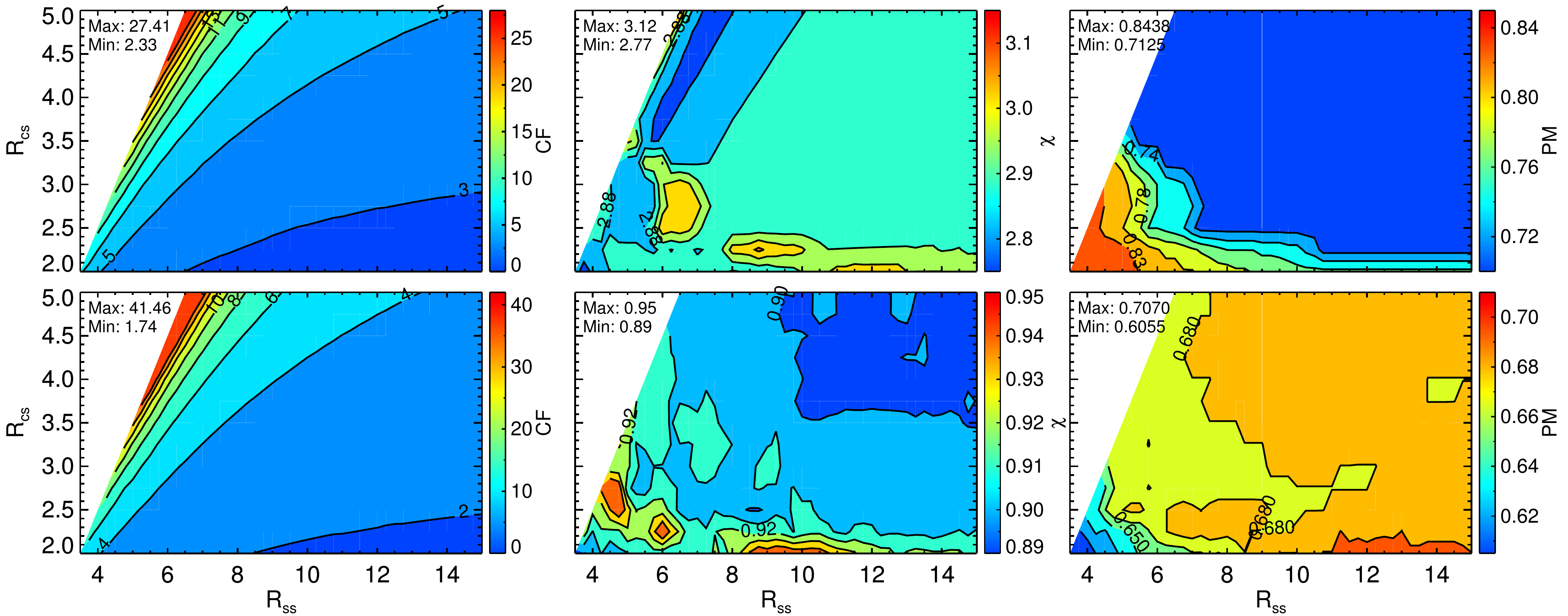}
\caption{Contour maps of the CF (left column), $\chi$ (middle column), and PM (right column), obtained by comparing the IMFs of the CSSS model with observations by SolO (top row) and OMNI (bottom row). The current parameter $a$ is set to $1.0\ R_\sun$. \label{fig:solo_omni_factor_pm}}
\end{center}
\end{figure*}

By comparing the predicted IMFs from the CSSS model with observations from SolO and OMNI, we show the contour maps of CF (left column), $\chi$ (middle column), and PM (right column) in Figure \ref{fig:solo_omni_factor_pm}. The current parameter $a$ is set to $1.0\ R_\sun$. Conclusions are consistent with the comparisons of the PSP. Specifically, when $R_{\rm cs}$ is small and $R_{\rm ss}$ is large, the CF attains optimal values. By examining the contour maps of $\chi$ and PM, it is evident that the optimal values require $R_{\rm cs}$ to not be excessively large. Therefore, it is suggested that for the CSSS model, the cusp surface $R_{\rm cs}$ should fall within the range of $2.0 - 2.4\ R_\sun$, while the source surface parameter $R_{\rm ss}$ should be within $11.0 - 15.0\ R_\sun$.

\section{Discussion and conclusions} \label{sec:conclusions}

\begin{figure*}[htb!]
\begin{center}
\includegraphics[width=1.0\textwidth]{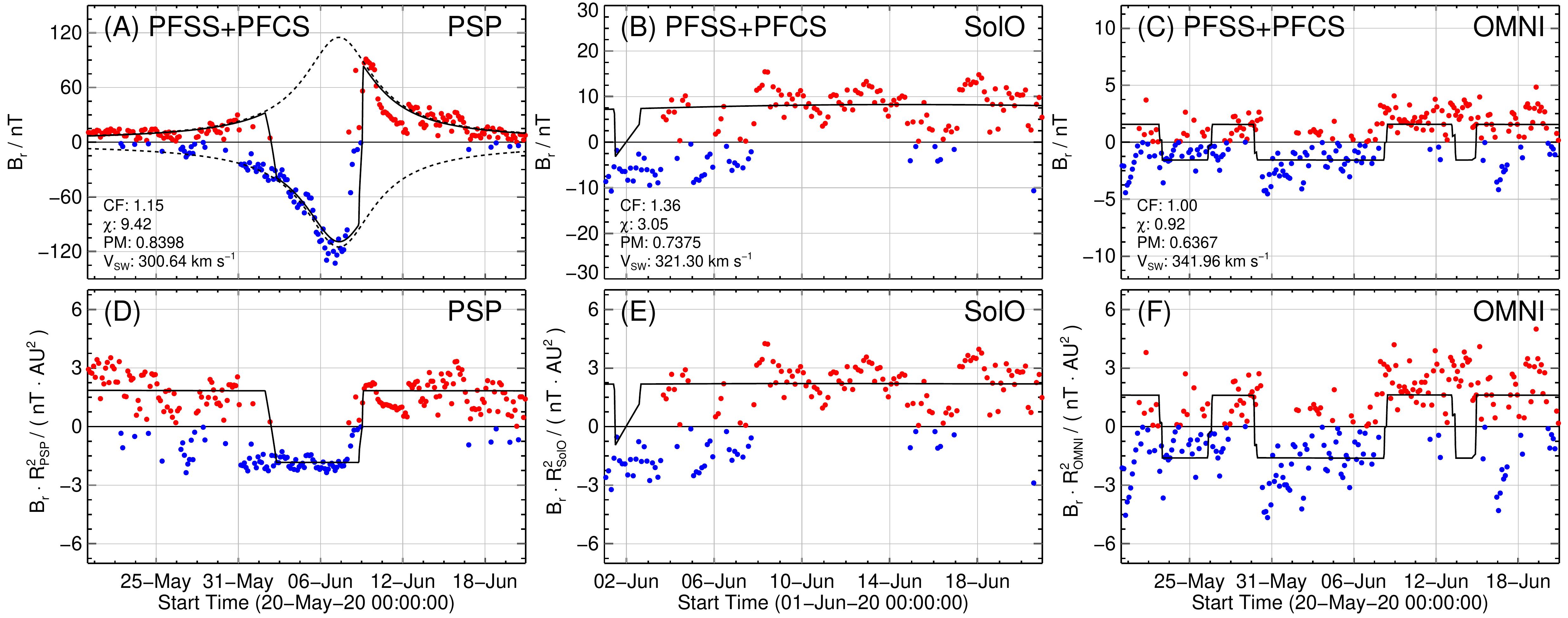}
\caption{Comparisons of the predicted IMFs by the PFSS+PFCS model with in-situ measurements (red and blue dots) of the PSP (A), SolO (B) and OMNI (C) using the HMI synoptic map introducing additional polar flux with a single Gaussian distribution. Bottom panels (D)-(F) show the quantity $B_rR^2$, indirectly reflecting the variations of the magnetic flux at different interplanetary positions. The free parameters of the PFSS+PFCS model are $R_{\rm ss} = 2.2\ R_\sun$ and $R_{\rm scs} = 12\ R_\sun$. \label{fig:in_situ_comp_add}}
\end{center}
\end{figure*}

In this study, we apply the PFSS+PFCS and CSSS models in conjunction with Parker spiral lines to examine the large-scale magnetic fields in both the corona and interplanetary space for CR 2231 during the solar activity minimum. Our analysis reveals that the magnetic field lines traced from the coronal models can effectively capture the observed structures such as streamers and plumes in the corona. Furthermore, the predictions of the IMFs are also consistent with the in-situ measurements when CFs are included. The primary objective of this research is to improve the global magnetic field extrapolated from the PFSS+PFCS and CSSS models constrained by remote-sensing and in-situ observations, while also optimizing free parameters in an attempt to comprehend the “missing” magnetic flux.

We calculate the magnetic field configuration derived from the PFSS+PFCS and CSSS models with various free parameters. We find that the HCSs in the PFSS+PFCS model present significant fluctuations when $R_{\rm ss}$ is smaller than $2.0\ R_\sun$. The magnetic field strength gradually decreases from high latitudes towards the HCSs when $R_{\rm scs}-R_{\rm ss}$ is smaller than $8.0\ R_\sun$. Additionally, the total unsigned magnetic flux $\phi_{\rm B}$ on the outer boundary decreases as $R_{\rm ss}$ or $R_{\rm scs}$ increases. Similarly, the CSSS model displays a substantial gradient in magnetic field strength $|B_r|$ near the HCSs when $R_{\rm ss}-R_{\rm cs}$ is smaller than $8.0\ R_\sun$. Notably, the topological configurations of the HCSs and the total unsigned magnetic flux $\phi_{\rm B}$ are primarily influenced by the cusp surface $R_{\rm cs}$ and the current parameter $a$, rather than the source surface $R_{\rm ss}$. From the profiles of the longitudinally averaged $|B_r|$, we find that increasing $R_{\rm scs}-R_{\rm ss}$ in the PFSS+PFCS model, and $R_{\rm ss}-R_{\rm cs}$ in the CSSS model, results in a closer magnetic field strength in the polar and equatorial regions. Figure \ref{fig:pfss_csss_mag_line} indicates that the locations of the horizontal current sheets obtained from the two types of models present good consistency.

We project the traced magnetic field lines derived from the PFSS+PFCS and CSSS models onto the PoS as observed from both Earth and SolO perspectives. We then compare these field lines with VL and EUV coronagraphic images. The magnetic field lines present a strong correlation with observed coronal structures, including streamers and plumes. The open magnetic field lines at the north and south poles of the Sun connect to the darker regions in the EUV image, while the closed magnetic field line structures are found within the streamers. The HCSs accurately capture the evolution of streamers in the Carrington maps.

Moreover, we predict the IMFs at locations of the PSP, SolO, and OMNI using the Parker spiral lines extrapolated from the outer boundaries of the PFSS+PFCS and CSSS models with different photospheric magnetograms. Three evaluation parameters, namely CF, $\chi$, and PM, have been established to assess magnetic field models. We find that the IMFs derived from the HMI synoptic map are better consistent in the strength of magnetic fields with in-situ measurements compared to the GONG daily synoptic maps. Furthermore, we find the CF is greatest at SolO, smallest at OMNI, and in between these values at PSP. As opposed to using a constant average, solar wind speed, adopting real-time measurements can only slightly improve the accuracy of the calculated IMFs. We conduct an extensive search for the optimal free parameters in PFSS+PFCS and CSSS models using the HMI synoptic map. We suggest that the optimal free parameter for the PFSS+PFCS model is $R_{\rm ss} = 2.2 - 2.5\ R_\sun$ and $R_{\rm scs} = 10.5 - 14.0\ R_\sun$. For the CSSS model, we suggest $R_{\rm cs} = 2.0 - 2.4\ R_\sun$ and $R_{\rm ss} = 11.0 - 14.7\ R_\sun$ for $a = 1.0\ R_\sun$.

Regardless of the adjustments made to the free parameters, the IMFs consistently showed an underestimation compared to the measurements. Therefore, we attempt to account for this underestimation by incorporating additional polar magnetic fields into the HMI map. The additional polar magnetic field is assumed to follow the single Gaussian distribution \citep{Riley2019a},
\begin{equation}
f(\theta') = B_{\rm{pol}}\cdot \rm{exp}(-\frac{\theta'^2}{2\sigma^2}),
\end{equation}
where $\theta'$ represents co-latitude in radians, and the standard deviation $\sigma$ is 0.25. The value of $\theta'$ is limited to a range of 0.25 rad from the north and south poles. The magnitude of the added magnetic field $B_{\rm{add}}$ is related to the peak strength $B_{\rm{pol}}$. We find that as the peak strength $B_{\rm{pol}}$ increases in the PFSS+PFCS and CSSS models, the HCSs on the source surfaces gradually flatten. This phenomenon can typically be attributed to the rise in magnetic pressure, which results in an additional transverse, equatorward pressure gradient. As shown in Figure \ref{fig:hmi_gong} (A), the total positive flux across the entire photosphere is $27.69 \times 10^{14}\ {\rm Wb}$ measured by the HMI, which balances the negative flux. To mitigate the underestimation of the calculated IMFs (i.e., achieving a CF close to 1) in the PFSS+PFCS model with $R_{\rm ss} = 2.2\ R_\sun$ and $R_{\rm scs} = 12\ R_\sun$, it is necessary to increase $B_{\rm{pol}}$ to $13.7\ {\rm G}$, which is equivalent to adding a total unsigned flux of $2.24 \times 10^{14}\ {\rm Wb}$ to each pole. After adding the additional polar magnetic field, the net positive or negative flux increases to $\pm 29.93 \times 10^{14}\ {\rm Wb}$. Panels (A)-(C) of Figure \ref{fig:in_situ_comp_add} present the comparisons of the PFSS+PFCS model between the IMFs obtained using HMI magnetic maps after artificially adding the polar magnetic fields and the in-situ measurements. In the CSSS model, there is a good agreement between the calculated IMFs and the measurements at $1\ {\rm AU}$ from the OMNI dataset when $B_{\rm{pol}} = 20.9\ {\rm G}$, resulting in a net positive/negative flux of $\pm 31.12 \times 10^{14}\ {\rm Wb}$. The free parameters of the CSSS model are $R_{\rm cs} = 2.2\ R_\sun$, $R_{\rm ss} = 12.0\ R_\sun$ and $a = 1.0\ R_\sun$. Nevertheless, from the comparison results of the PSP and the SolO, the OFP appears to persist. The bottom panels of Figure \ref{fig:in_situ_comp_add} present the variations of $B_rR^2$ at the positions of the PSP (panel D), SolO (panel E), and Earth (panel F), indirectly indicating that the measured magnetic flux is not conserved. The variations of $B_rR^2$ could be related to local solar wind properties influenced by the complex interplanetary processes, such as CMEs, shocks, stream interfaces, and switchbacks. Future research endeavors will concentrate on investigating the magnetic activities within interplanetary space, including the dynamic interactions between particles and fields.

\begin{acknowledgments}

We sincerely thank the anonymous referee for providing valuable suggestions that helped us to improve the quality of the manuscript. We thank Jie Jiang for helpful discussions on the CSSS model. We also thank the teams of SDO/HMI, GONG, K-Cor, SOHO/LASCO, PSP/FIELDS, PSP/SWEAP, SolO/MAG, SolO/EUI for their open-data use policy. SDO and PSP are missions of NASA's Living with a Star (LWS) Program. GONG is a community-based program, managed by the National Solar Observatory (NSO). The K-Cor data are courtesy of the MLSO, operated by the High Altitude Observatory (HAO), as part of the National Center for Atmospheric Research (NCAR). SOHO and SolO are missions of international cooperation between ESA and NASA. The authors also thank NASA/GSFC's Space Physics Data Facility's OMNIWeb service (\url{https://omniweb.gsfc.nasa.gov/}), and are thankful to the OMNIWeb team. This work is supported by the Strategic Priority Research Program of the Chinese Academy of Sciences, Grant No. XDB0560000, National Key R\&D Program of China 2022YFF0503003 (2022YFF0503000), NSFC (Grant No. 11973012, 11921003, 12103090, 12203102, 12233012), the mobility program (M-0068) of the Sino-German Science Center. This work benefits from the discussions of the ISSI-BJ Team “Solar eruptions: preparing for the next generation multi-waveband coronagraphs”, and is also inspired by the sun shadow data measured by LHAASO.

\end{acknowledgments}

%



\software{Astropy \citep{Astropy2013, Astropy2018, Astropy2022}, Sunpy \citep{sunpy2020}, Pfsspy \citep{Yeates2018,Stansby2020}, POT3D \citep{Caplan2021ApJ}}





\bibliographystyle{aasjournal}
\bibliography{glshi_biblio}{}



\end{document}